\documentstyle[aps,preprint,latexsym,amscd,amsmath]{revtex}
\newcounter{saveeqn}%
\newcommand{\alphaeqn}{\setcounter{saveeqn}{\value{equation}}%
\stepcounter{saveeqn}\setcounter{equation}{0}%
\renewcommand{\theequation}
        {\mbox{\arabic{saveeqn}-\alph{equation}}}}%
\newcommand{\reseteqn}{\setcounter{equation}{\value{saveeqn}}%
\renewcommand{\theequation}{\arabic{equation}}}%

\begin{document}
\title{Symmetries and conservation laws in histories-based theories}
\author{Tulsi Dass\footnote{Electronic mail: tulsi@iitk.ac.in}  and Yogesh N. Joglekar\footnote{Electronic mail: yojoglek@indiana.edu}\thanks{Present address: Department of Physics, Indiana University, Bloomington, IN 47405} \\ Department of Physics, \\ Indian Institute of Technology, Kanpur, India 208016}
\maketitle


\begin{abstract}
Symmetries are defined in histories-based theories paying special attention to the class of history theories admitting quasitemporal structure (a generalization of the concept of `temporal sequences' of `events' using partial semigroups) and logic structure for `single-time histories'. Symmetries are classified into orthochronous (those preserving the `temporal order' of `events') and nonorthochronous. A straightforward criterion for physical equivalence of histories is formulated in terms of orthochronous symmetries; this criterion covers various notions of physical equivalence of histories considered by Gell-Mann and Hartle as special cases. In familiar situations, a reciprocal relationship between traditional symmetries (Wigner symmetries in quantum mechanics and Borel-measurable transformations of phase space in classical mechanics) and symmetries defined in this work is established. In a restricted class of theories, a definition of conservation law is given in the history language which agrees with the standard ones in familiar situations; in a smaller subclass of theories, a Noether type theorem (implying a connection between continuous symmetries of dynamics and conservation laws) is proved. The formalism evolved is applied to histories (of particles, fields or more general objects) in general curved spacetimes. Sharpening the definition of symmetry so as to include a continuity requirement, it is shown that a symmetry in our formalism implies a conformal isometry of the spacetime metric.
\end{abstract}

\tableofcontents


\section{Introduction}
\label{sec: intro}

Back in the mid-sixties, Houtappel, Van Dam, and Wigner~\cite{hvw1} presented a general treatment of geometric invariance principles (\emph{i.e.} those invariance principles which correspond to transformations between equivalent reference frames) in classical and quantum mechanics in terms of the primitive elements of a physical theory. These primitive elements were observations or measurements and their results. To describe the correlations between observations, they employed the conditional probabilities $\Pi(A|B)$ where $A=(\alpha,r_\alpha;\beta,r_\beta,\cdots;\epsilon,r_\epsilon)$ represented a set of measurements $\alpha,\beta,\ldots,\epsilon$ (at times $t_\alpha,t_\beta,\cdots,t_\epsilon$) with respective outcomes $r_\alpha,r_\beta,\cdots,r_\epsilon$ and similarly $B=(\zeta,r_\zeta;\eta,r_\eta;\ldots;\nu,r_\nu)$; the quantity $\Pi(B|A)$ represented the probability of realization of $B$, given $A$. (We have changed the notation in~\cite{hvw1} from $\Pi(A|B)$ to  $\Pi(B|A)$ to bring it in correspondence with standard usage in probability theory.) The ordering of times $(t_\alpha,t_\beta,\cdots,t_\epsilon,t_\zeta,t_\eta,\ldots,t_\nu)$ was arbitrary and all measurements referred to external observers. The conditional probabilities $\Pi(B|A)$ are quite general and can be employed in classical as well as quantum mechanics.

The idea of addressing fundamental questions in physics in a general formalism employing only primitive elements of physical theory is an attractive one. In the past one and a half decade, it has been pursued employing `histories' which are closely related to the $\Pi$-functions. In the history version of quantum theory, a history of a system S is a time-ordered sequence of `events'
\begin{equation}
\label{eq: intro1}  		
\alpha = (\alpha_{t_1},\alpha_{t_2},\ldots,\alpha_{t_n}) \hspace{1in} t_1<t_2<\ldots<t_n 
\end{equation}				
where $\alpha_{t_i}$ are the Schr\"{o}dinger picture projection operators. In traditional quantum mechanics [assuming, as usual, that the projections in Eq. (\ref{eq: intro1}) represent measurements by external observers] the probability of the history (\ref{eq: intro1}) is given by 
\begin{equation}
\label{eq: intro2}	       
P(\alpha) = \mbox{Tr}\left[ C_\alpha \rho(t_0) C_\alpha^\dagger \right]
\end{equation}				
where
\begin{equation}
\label{eq: intro3}		
C_\alpha = \alpha_{t_n} U(t_n,t_{n-1}) \alpha_{t_{n-1}} \ldots \alpha_{t_2}U(t_2,t_1) \alpha_{t_1} U(t_1,t_0),
\end{equation}			
 $U(t,t') = \exp\left[-iH(t-t')/\hbar\right] $ is the evolution operator and $\rho(t_0)$ is the initial density operator at a time $t_0<t_1$.

 Griffiths and Omnes~\cite{rbg1,ro1,ro2,ro3,go1} have proposed an interpretive scheme for traditional quantum mechanics in which the Hilbert space based mathematical formalism is retained, the reduction postulate is discarded and Eq. (\ref{eq: intro2}) is interpreted as the probability for the history (\ref{eq: intro1}) for a \emph{closed} system S (no external observers). As all the probabilities employed are classical, the probability assignment can be made only for histories satisfying appropriate `consistency conditions' (or `decoherence conditions') ensuring the absence of quantum mechanical interference in the relevant family of histories. For certain pairs of histories $\alpha,\beta$ with $\alpha$ as in Eq. (\ref{eq: intro1}) and $\beta = (\beta_{t_1},\ldots,\beta_{t_m})$ with the same initial state $\rho(t_0)$, the decoherence condition takes the form 
\begin{equation}
\label{eq: intro4}		
\mbox{Re}[d(\alpha,\beta)] = 0,
\end{equation}				
where the so-called \emph{decoherence functional} $d(\alpha,\beta)$ is given by
\begin{equation}
\label{eq: intro5}		
d(\alpha,\beta) = \mbox{Tr} [C_\alpha \rho(t_0) C_\beta^\dagger].
\end{equation}				
Note that
\begin{equation}
\label{eq: intro6}			
 P(\alpha) = d(\alpha,\alpha). 
\end{equation}				

In histories-based theories, one takes, following the ideas of Gell-Mann and Hartle~\cite{gh1,jh1,jh2}, histories as the basic objects. In these theories, the time sequences employed in the description of histories like (\ref{eq: intro1}) serve only for book-keeping; the properties of time $t$ as a real variable are not used. The mathematical structure which correctly describes the book-keeping in histories-based theories and also makes provision for useful generalizations of the concept of time is that of a  partial semigroup~\cite{cji1}.

A \emph{partial semigroup} (psg) is a nonempty set ${\mathcal K}$ (whose elements will be denoted as $s,t,u,\ldots$) in which a binary operation $\circ$ between certain pairs of elements is defined such that $(s\circ t)\circ u = s\circ(t\circ u)$ whenever both sides are well-defined. A homomorphism of a psg ${\mathcal K}$ into another psg ${\mathcal K}'$ is a mapping $\sigma:{\mathcal K}\rightarrow{\mathcal K}'$ such that, for all  $s,t\in{\mathcal K}$ with $s\circ t$ defined, $\sigma(s)\circ\sigma(t)$ is also defined and 
\begin{equation}
\label{eq: intro7}		
\sigma(s \circ t) = \sigma(s)\circ\sigma(t).
\end{equation}				
If $\sigma$ is invertible, it is called an isomorphism (automorphism if ${\mathcal K}'={\mathcal K}$). The terms antihomomorphism, anti-isomorphism and antiautomorphism are similarly defined with the order of terms on the right in Eq. (\ref{eq: intro7}) reversed.

The partial semigroups involved in the book-keeping of histories in quantum mechanics are ${\mathcal K}_1$ and ${\mathcal K}_2$ defined as follows. We have
\begin{displaymath}			
{\mathcal K}_1 = \{\mbox{finite ordered subsets of \emph{R}}\}.
\end{displaymath}			        
A general element $t\in{\mathcal K}_1$ is of the form 
\begin{equation}
\label{eq: intro8}		
t=\{t_1,t_2, \ldots, t_n\}; \hspace{1in} t_1<t_2<\ldots<t_n.
\end{equation}				
If $s =\{s_1,s_2,\ldots,s_m\}\in{\cal K}_1$ such that $s_m<t_1$, then $s\circ t$ is defined and 
\begin{equation}
\label{eq: intro9}
s\circ t = \{s_1,s_2,\ldots,s_m,t_1,t_2,\ldots,t_n\}.
\end{equation}				
We adopt the convention~\cite{dj1}
\begin{equation}
\label{eq: intro10}		
\{t_1\}\circ\{t_1\} = \{t_1\}.
\end{equation}				
 With this convention, we have $s\circ t$ defined for $s_m\leq t_1$. Note that elements of ${\mathcal K}_1$ admit irreducible decomposition of the form
\begin{equation}
\label{eq: intro11}	
t = \{t_1\}\circ \{t_2\}\circ \cdots \circ \{t_n\}.
\end{equation}			
Elements $\{t_i\}$ which cannot be further decomposed are called \emph{nuclear}.

The other psg ${\mathcal K}_2$ consists of histories as its elements. For $\alpha = \{\alpha_{s_1},\alpha_{s_2},\ldots,\alpha_{s_m}\}$ and $\beta = \{\beta_{t_1},\beta_{t_2},\ldots,\beta_{t_n}\}$ with $s_m< t_1$, $\alpha\circ\beta$ is defined and is given by
\begin{equation}
\label{eq: intro12}		
\alpha\circ\beta = \{\alpha_{s_1},\alpha_{s_2},\ldots,\alpha_{s_m},\beta_{t_1},\beta_{t_2},\ldots,\beta_{t_n}\}.		
\end{equation}				
There is a homomorphism $\sigma$ from ${\mathcal K}_2$ onto ${\mathcal K}_1$ given by
\begin{eqnarray}
\label{eq: intro13}		
\sigma(\alpha) = s,\hspace{1cm} & \sigma(\beta)= t,\hspace{1cm} & \sigma(\alpha\circ\beta) = s\circ t.
\end{eqnarray}				
The triple $({\mathcal K}_2,{\mathcal K}_1, \sigma)$ defines a \emph{quasitemporal structure} (a pair of psg's with a homomorphism of one onto the other). Note that, given a single time element $\{t_1\}\in{\mathcal K}_1$, the space $\left({\mathcal K}_2\right)_{t_1} = \sigma^{-1}\{t_1\}$ is the set ${\mathcal P}({\mathcal H})$ of projection operators in the quantum mechanical Hilbert space ${\mathcal H}$ of the system; in the framework of quantum logic~\cite{jmj1,vsv1}, these projection operators represent single-time propositions. The space ${\mathcal P}({\mathcal H})$ constitutes a \emph{logic} in the sense of Varadarajan~\cite{vsv1}.

The concept of quasitemporal structure is a generalization of the idea of histories as temporal sequences of `events'. With suitably chosen psg's (employing light cones etc.) this concept serves to provide a framework general enough to accommodate history versions of quantum field theories in curved space-times~\cite{cji1}.

Taking clue from the traditional proposition calculus where single-time propositions are the basic entities, Isham suggested that histories must be treated as (multi-time or more general) propositions. He evolved a scheme of `quasitemporal theories' in which the basic objects were a triple $({\mathcal U},{\mathcal T},\sigma)$ defining a quasitemporal structure. The space ${\mathcal U}$ was called the `space of history filters' and was assumed to be a meet semilattice with the operations of partial order $\leq$ (coarse graining) and a meet operation $\wedge$ (simultaneous realization of two histories). The space ${\mathcal T}$ was called the `space of temporal supports'. To accommodate the operation of negation of a history, he proposed that the space ${\mathcal U}$ be embedded in a larger space $\Omega$ called the `space of history propositions' (denoted as ${\mathcal {UP}}$ in~\cite{cji1} and~\cite{il1}). This larger space was envisaged as having a lattice structure. Decoherence conditions and probabilities of decoherent histories were to be defined in terms of decoherence functionals [see Eqs. (\ref{eq: intro4}) and (\ref{eq: intro6}) above] which were complex valued functions satisfying the standard four conditions of hermiticity, positivity, bilinearity and normalization. [See Eqs. (\ref{eq: atlf9a})-(\ref{eq: atlf11}) below.]  
						
A more general scheme was later proposed by Isham and Linden~\cite{il1} in which the basic objects were the two spaces $(\Omega,{\mathcal D})$ where $\Omega$ (the space of history propositions) was assumed to be an orthoalgebra incorporating a partial order $\leq$ (coarse graining), a disjointness relation $\bot$ (mutual exclusion), a join operation $\oplus$ (`or' operation for mutually exclusive propositions), and a few other features. The space ${\mathcal D}$ was the space of decoherence functionals satisfying the above mentioned properties. The quasitemporal theories are a subclass of this general class of theories.

In a recent paper~\cite{dj1} we have presented axiomatic development of dynamics of  systems in the framework of histories which contains the history versions of classical and traditional quantum mechanics as special cases. We considered theories which admit quasitemporal structure $({\mathcal U},{\mathcal T},\sigma)$ in the sense explained above.  We assumed that the spaces ${\mathcal U}_\tau = \sigma^{-1}(\{\tau\})$ for nuclear elements $\{\tau\}\in{\mathcal T}$ (the spaces of single time propositions) have the structure of a logic. Isomorphism of ${\mathcal U}_\tau$'s (as logics) at different $\tau$'s was not assumed. This generality gives the formalism additional flexibility so as to make it applicable to systems whose empirical characteristics may change with time; for example, the Universe. Using the logic structure of ${\mathcal U}_\tau$'s, a larger space $\Omega$ - the space of history propositions was explicitly constructed and shown to be an orthoalgebra as envisaged in the scheme of Isham and Linden; its subspace $\tilde{\mathcal U}$ representing `homogeneous histories' (which is obtained from ${\mathcal U}$ after removing some redundancies) was shown to be a meet semilattice as envisaged in the scheme of Isham~\cite{cji1}. Explicit expressions for decoherence functionals were given for the Hilbert space based theories with ${\mathcal U}_\tau = {\mathcal P}({\mathcal H}_\tau)$ (the lattice of projection operators in a separable Hilbert space ${\mathcal H}_\tau$) and for classical mechanics. 

The present work is devoted to a systematic treatment of symmetries and conservation laws in histories-based theories. We choose the above mentioned formalism for a detailed treatment of symmetries since the mathematical structure of this formalism facilitates treatment of some detailed features of symmetry operations in history theories. Following the general idea of defining symmetries as structure-preserving invertible mappings in appropriate mathematical framework~\cite{hw1}, we define a symmetry as a triple $\Phi=(\Phi_1,\Phi_2,\Phi_3)$ of invertible mappings ($\Phi_1:{\mathcal T}\rightarrow{\mathcal T},\Phi_2:{\mathcal U}\rightarrow{\mathcal U}$, and $\Phi_3:{\mathcal D}\rightarrow{\mathcal D}$) preserving the quasitemporal structure, the logic structure of single `time' propositions, the decoherence condition and the probabilities of histories (section~\ref{sec: sihbgqm}). A natural classification of symmetries as \emph{orthochronous} and \emph{non-orthochronous} appears. 

From our definition of symmetry in the formalism of~\cite{dj1} it is clear how symmetries are to be defined in general quasitemporal theories and, more generally, in the Isham and Linden scheme. A good example in this context is the work of Schreckenberg~\cite{ss1} who has considered symmetries in the subclass of Isham-Linden type theories treated by Isham, Linden, and Schreckenberg~\cite{ils1}.

 In sections~\ref{sec: sinqm} and \ref{sec: sihvocm} we consider the (history versions of) traditional quantum mechanics and classical mechanics respectively. In these cases, we establish a reciprocal relationship between symmetries defined here and symmetries of the formalism of traditional sort - Wigner symmetries in quantum mechanics and Borel measurable transformations of the phase space in classical mechanics (the latter are the natural structure preserving transformations in classical mechanics in the framework of logics). 

Gell-Mann and Hartle~\cite{gh3} have considered several notions of physical equivalence of histories in the quantum mechanics of closed systems. Using the concept of orthochronous symmetries we obtain an economic formulation of the concept of physical equivalence of histories which covers all these notions as special cases (section~\ref{sec: peoh}).

It is of some interest to treat conservation laws in the framework of histories. To admit the possibility of defining conservation laws, the formalism must permit comparison of observations at different `times'; it follows that conservation laws can be defined only if the spaces of single `time' propositions, ${\mathcal U}_\tau$'s at various instants of `time' are mutually isomorphic. We first give, in section \ref{sec: cs}, a straightforward definition of conservation law in terms of temporal evolution described by mappings between pairs of ${\mathcal U}_\tau$'s. It is then translated into a (supposedly equivalent) definition in terms of decoherence functionals. The equivalence of the two definitions is verified in (history versions of) traditional quantum mechanics and classical mechanics. We do not consider the deeper question of the conservation of an observable in histories involving `events' relating to other observables as well and the constraints implied by such conservation. Such questions have been considered by Hartle {\it et al.}~\cite{hlm1} in the context of the Hilbert space quantum mechanics (of closed systems). That work employs a definition of conservation law somewhat different from ours; this is briefly discussed in section~\ref{sec: cs}.

In any scheme of mechanics, one expects a general connection between continuous symmetries of dynamics and conservation laws. The most famous result of this type is the Noether's theorem in Lagrangian dynamics~\cite{elh1,td1}. In our formalism, the twin requirements of an explicit expression for the decoherence functional and interpretation of the infinitesimal generators of a symmetry in terms of observables restricts the possibility of proving such a theorem only for the Hilbert space based theories in which the Hilbert spaces ${\mathcal H}_\tau$ corresponding to various nuclear $\tau$ are naturally isomorphic. A theorem showing that, in such theories, a continuous symmetry implies a conservation law is proved in section~\ref{sec: cbsacl}. The commutation of the evolution operator and the generator of symmetry transformation, which is used in the above theorem, is shown in appendix~\ref{app: two}.

In section~\ref{sec: hics}, we consider, in some detail, histories (of particles, fields or more general objects) in general curved spacetimes. In spacetimes permitting foliation in spacelike hypersurfaces, the parameter labeling the leaves of foliation plays the role of time. Discrete values of the same parameter can be employed for book-keeping of histories~\cite{mb1}. Basic concepts relating to the treatment of histories of quantum fields in general spacetimes were developed by Hartle~\cite{jh1} and Sorkin~\cite{rds1} who employed averages of fields over spacetime regions as basic observables. A temporal order was defined on a causally consistent family of spacetime regions ({\it i. e.} a family having no pair of regions in which one region intersects both the future and the past of the other). Appropriate families of such regions served as `time points' in the description of field histories. Isham translated these ideas in the language of partial semigroups leading to a quasitemporal structure $({\mathcal U},{\mathcal T},\sigma)$ for such theories. Given a spacetime $(M,g)$, he defined a temporal support as a collection of (four dimensional) `basic regions' with appropriate temporal relation. This involves a relation, $\prec$, defined as follows: Given subsets $A$ and $B$ of $M$,  $A\prec B$ ($A$ `temporally precedes' $B$) if 
\begin{eqnarray}
\label{eq: intro14}
A\cap B=\phi,\hspace{1cm} & J^+(A)\cap B\neq\phi,\hspace{1cm} & J^+(B)\cap A=\phi.
\end{eqnarray}
Here $J^+(A)$ is the causal future of $A$ (\emph{i. e.} the set of points in $M$ that can be reached from $A$ by future directed non-spacelike curves).

We note that the temporal supports employed by Blencowe [finite ordered sequences of (subsets of) spacelike hypersurfaces] cannot be considered as a subclass of the temporal supports as defined by Isham because the latter employ four-dimensional basic regions. From a theoretical point of view, the use of four-dimensional regions is more satisfying because it makes provision for finite spatio-temporal localization of `events'; however it is quite often convenient and useful to work with idealized `time points' (which means, in the present context, three or lower dimensional basic regions). In section~\ref{sec: hics} we give the construction of the space of temporal supports employing such idealized `time points'; it can be used to generalize Blencowe's treatment of field histories to general spacetimes. The quasitemporal structure is chosen so as to conform to the conventions of~\cite{dj1}. This necessitates, apart from the replacement of four dimensional basic regions by three (and lower) dimensional basic regions, a modification of the temporal order relation (\ref{eq: intro14}) of Isham [see Eq. (\ref{eq: qs1}) below]. 

In the treatment of symmetries, we impose the condition of continuity on the mappings mentioned above; this is done by taking the spaces ${\mathcal U}$ and ${\mathcal T}$ to be topological partial semigroups. We concentrate on the mapping $\Phi_1:{\mathcal T}\rightarrow{\mathcal T}$ (which is a part of definition of a symmetry), construct a topology for ${\mathcal T}$ explicitly (appendix~\ref{app: one}) and show that, with the continuity requirement imposed, the mapping $\Phi_1$ induces a transformation of the spacetime $M$ which is a conformal isometry of the metric $g$. 

The last section on concluding remarks includes, among others, a detailed comment on the histories-based theories characterized as theories describing dynamics of systems in terms of primitive elements of physical theory.


\section{ Histories-based generalized quantum mechanics}
\label{sec: hbgqm}

In this section, we shall recapitulate some essential points from~\cite{dj1}.


\subsection{More about partial semigroups}
\label{subsec: maps}

A \emph{unit element} in a psg ${\mathcal K}$ is an element $e$ such that $e\circ s = s\circ e = s$ for all $s\in{\mathcal K}$. An \emph{absorbing element} in ${\mathcal K}$ is an element $a$ such that for all $s\in{\mathcal K}$, $a\circ s = s\circ a = a $. A psg may or may not have a unit and/or absorbing element; when either of them exists, it is unique. In a psg, elements other than the unit and the absorbing elements are called \emph{typical}. If a psg ${\mathcal K}$ has a unit element $e$ and/or an absorbing element $a$ and if there is a homomorphism $\sigma$ from ${\mathcal K}$ onto ${\mathcal K}'$, then  ${\mathcal K}'$ must correspondingly have a unit element $e'$ and/or an absorbing element $a'$ such that 
\begin{equation}
\label{eq: psg1}	
\sigma(e) = e', \hspace{.3in} \sigma(a) = a'.
\end{equation}
We call a psg ${\mathcal K}$ \emph{directed} if, for any two different typical elements $s, t$ in ${\mathcal K}$, when $s\circ t$ is defined then $t\circ s$ is not defined. The psg's ${\mathcal K}_1$ and ${\mathcal K}_2$ introduced in section~\ref{sec: intro} are directed. In a theory with a quasitemporal structure, the concept of direction of flow of `time' can be introduced by taking the two psg's to be directed.

The concept of `point of time' gets replaced in the psg setting by that of a nuclear element. The nuclear elements of ${\mathcal K}_1$ are, indeed, points of time. [See Eq.(\ref{eq: intro11}).] In ${\mathcal K}_2$, the nuclear elements are of the form $\{\alpha_t\}$ representing single time history propositions. The set of nuclear elements in a psg ${\mathcal K}$ is denoted as ${\mathcal N}({\mathcal K})$. Clearly ${\mathcal N}({\mathcal K}_1) = R $, the set of real numbers.

A psg is called \emph{special} if its typical elements admit semi-infinite irreducible decompositions [modulo redundancies implied by the convention (\ref{eq: intro10})]. The psg's ${\mathcal K}_1$ and ${\mathcal K}_2$ are the trivial examples of special psg's. A nontrivial example is the psg ${\mathcal K}_3$ whose elements are ordered subsets of $R$ which are at most countably semi-infinite (with elements of the form $s=\{s_1,\ldots,s_n\}, s'=\{s_1,s_2,\ldots\}, s''=\{\ldots,s_{-1},s_0\}$) with composition rule a straightforward extension of that of ${\mathcal K}_1$~\cite{dj1}.


\subsection{The augmented temporal logic formalism}
\label{subsec: atlf}

The axiomatic development of generalized quantum mechanics of closed systems~\cite{dj1} is structured around five axioms $A_1,\ldots,A_5$.

\begin{enumerate}
\item[]\noindent\emph{$A_1$: (Quasitemporal Structure Axiom)}: 
Associated with every dynamical system is a history system $({\mathcal U},{\mathcal T},\sigma)$ defining a quasitemporal structure as described in section~\ref{sec: intro}. The psg's ${\mathcal U}$ and ${\mathcal T}$ are assumed to be special and satisfy the relation 
\begin{equation}
\label{eq: atlf1}
\sigma\left[{\mathcal N}({\mathcal U})\right] = {\mathcal N}({\mathcal T}).
\end{equation}				
\end{enumerate}
Elements of ${\mathcal U}$ (history filters) are denoted as $\alpha,\beta,\ldots$ and those of ${\mathcal T}$ (temporal supports) as $\tau,\tau',\ldots$. If $\alpha\circ\beta$ and $\tau\circ\tau'$ are defined, we write $\alpha\lhd \beta$ ($\alpha$ precedes $\beta$) and $\tau\lhd\tau'$ ($\tau$ precedes $\tau'$).

\begin{enumerate}
\item[] {\bfseries \emph{$A_2$: (Causality Axiom)}}: If $\alpha,\beta,\ldots,\gamma\in{\mathcal N}({\mathcal U})$ are such that $\alpha\lhd\beta\lhd\ldots\lhd\gamma$ with $\sigma(\alpha) = \sigma(\gamma)$ then we must have $\alpha = \beta = \ldots = \gamma$.
\end{enumerate}
In essence this axiom forbids histories corresponding to `closed time loops'. From these two axioms we can prove that the psg's ${\mathcal U}$ and ${\mathcal T}$ are directed and some other useful results.

\begin{enumerate}
\item[] {\bfseries\emph{$A_3$: (Logic Structure Axiom)}}: Every space ${\mathcal U}_\tau = \sigma^{-1}(\tau)$ for $\tau\in{\mathcal N}({\mathcal T})$ has the structure of a logic as defined in Varadarajan~\cite{vsv1}.
\end{enumerate}


The remaining two axioms will be stated after outlining a few constructions using the above axioms.

We do not assume isomorphism of ${\mathcal U}_\tau$'s (as logics) for different $\tau$'s. Every ${\mathcal U}_\tau$ has two distinguished elements $0_\tau$ (the null proposition) and $1_\tau$ (the unit proposition) such that $0_\tau\leq\alpha\leq 1_\tau$ for all $\alpha\in{\mathcal U}_\tau$. For a typical  $\alpha\in{\mathcal U}$, we define $supp(\alpha)$ (called the temporal support of $\alpha$) as the unique collection of elements of ${\mathcal N}({\mathcal T})$ appearing in the irreducible decomposition of $\sigma(\alpha)$. For any $\tau\in supp(\alpha)$, the nuclear element in the irreducible decomposition of $\alpha$ projecting onto $\tau$ under $\sigma$ is denoted as $\alpha_\tau$.

The possible presence of $0_\tau$'s and $1_\tau$'s in irreducible decompositions causes some redundancy which needs to be removed. We call an $\alpha\in{\mathcal U}$ a \emph{null history filter} if $\alpha_\tau = 0_\tau$ for at least one $\tau\in supp(\alpha)$; we call it a \emph{unit history filter} if $\alpha_\tau = 1_\tau$ for all $\tau\in supp(\alpha)$. All the null history filters are physically equivalent (they represent absurd histories) as are all unit history filters (which represent the trivial histories involving propositions which are always true). We remove the redundancy by introducing an equivalence relation in ${\mathcal U}$ such that all null history filters are treated as equivalent and so are all non-null filters with the same reduced form (the form obtained by deleting the redundant $1_\tau$'s in the irreducible decomposition). We denote the equivalence class of $\alpha\in{\mathcal U}$ by $\tilde{\alpha}$. The set $\tilde{\mathcal U}$ of the equivalence classes in ${\mathcal U}$ inherits a psg structure from ${\mathcal U}$. In this psg, the equivalence class $\tilde{0}$ of all null history filters acts as an absorbing element and the equivalence class $\tilde{1}$ of all unit history filters acts as the unit element. 
								
A psg $\tilde{\mathcal T}$ is constructed from ${\mathcal T}$ in a similar fashion. The typical elements of $\tilde{\mathcal T}$ and the homomorphism $\tilde{\sigma}:\tilde{\mathcal U}\rightarrow\tilde{\mathcal T}$ (restricted to typical elements of $\tilde{\mathcal U}$) is obtained by defining $\tilde{\tau} = \tilde{\sigma}(\tilde{\alpha})$ to be the reduced temporal support obtained from $\tau = \sigma(\alpha)$ (for any representative $\alpha$ of $\tilde{\alpha}$) by deleting those elements which are images under $\sigma$ of the redundant $1_\tau$'s in the irreducible decomposition of $\alpha$. The unit element $\tilde{e}$ and the absorbing element $\tilde{a}$ are defined as
\begin{equation}
\label{eq: atlf2}
\tilde{e} = \phi\hspace{.5cm}[\mbox{the empty subset of }{\mathcal N}({\mathcal T})], \hspace{0.3in} \tilde{a} = {\mathcal N}({\mathcal T})	
\end{equation}		
and the homomorphism $\tilde{\sigma}$ is extended to include the relations 
\begin{equation}
\label{eq: atlf3}
\tilde{\sigma}(\tilde{1}) = \tilde{e}, \hspace{0.3in} \tilde{\sigma}(\tilde{0}) = \tilde{a} .
\end{equation}				
Defining, for any $\tilde{\alpha}\in\tilde{\mathcal U}, supp(\tilde{\alpha})=\tilde{\sigma}(\tilde{\alpha})$ [considered as a subset of ${\mathcal N}(\tilde{\mathcal T}) = {\mathcal N}({\mathcal T})$], we have
\begin{equation}
\label{eq: atlf4}
supp(\tilde{1})=\phi, \hspace{1in} supp(\tilde{0})={\mathcal N}({\mathcal T}).
\end{equation}
 
Using irreducible decompositions and the logic structure on ${\mathcal U}_\tau$'s one can define (in an intuitively suggestive manner) partial order ($\leq$), disjointness ($\bot$), disjoint join operation ($\oplus$), and meet operation ($\wedge$) in $\tilde{\mathcal U}$. The structure $(\tilde{\mathcal U},\leq,\wedge)$ is a meet semilattice as envisaged in Isham's scheme~\cite{dj1}. It is the triple $(\tilde{\mathcal U},\tilde{\mathcal T},\tilde{\sigma})$ [and not the original $({\mathcal U},{\mathcal T},\sigma)$] which corresponds to the triple in~\cite{cji1}.

The embedding of  $\tilde{\mathcal U}$ in a larger space $\Omega$ of history propositions is realized as follows. We define $\Omega$ to be the space of at most countable collections of mutually orthogonal elements of  $\tilde{\mathcal U}$ such that the union of temporal supports of any finite subcollection of them is orientable. (A subset $A$ of a psg ${\mathcal K}$ is said to be \emph{orientable} if it is at most countable and if there exists an ordering of elements of $A$ such that composition of every pair of consecutive elements is defined.) We denote elements of $\Omega$ as $\underline{\alpha},\underline{\beta},\ldots$. The space $\Omega$ has a null element $\underline{0} = \{\tilde{0}\}$ and a unit element $\underline{1} = \{\tilde{1}\}$. Elements of $\Omega$ other than $\underline{0}$ and $\underline{1}$ are called \emph{generic}.

In $\Omega$, we can define the operations of partial order ($\leq$), disjointness ($\bot$), disjoint join operation ($\oplus$) and show that, with these operations, it is an orthoalgebra (as envisaged in the scheme of Isham and Linden). A general element $\underline{\alpha} = \{\tilde{\alpha}^{(1)},\tilde{\alpha}^{(2)},\ldots\}$ of $\Omega$ can also be represented as 
\begin{equation}
\label{eq: atlf5}
\underline{\alpha} = \{\tilde{\alpha}^{(1)}\}\oplus\{\tilde{\alpha}^{(2)}\}\oplus\cdots
\end{equation}
A collection $\{\underline{\alpha},\underline{\beta},\ldots\}$ of mutually disjoint elements of $\Omega$ is said to be \emph{complete} (or \emph{exhaustive}) if
\begin{equation}
\label{eq: atlf6}		
\underline{\alpha}\oplus\underline{\beta}\oplus\cdots = \underline{1} .
\end{equation}				

The logic structure of ${\mathcal U}_\tau$'s permits us to introduce a space ${\mathcal S}(\tau)$ of states and a space ${\mathcal O}(\tau)$ of observables at `time' $\tau$ as in the traditional proposition calculus. A state at `time' $\tau\in{\mathcal N}({\mathcal T})$ is a generalized probability on ${\mathcal U}_\tau$, \emph{i.e.} a map $p_\tau:{\mathcal U}_\tau\rightarrow R$ such that (i) $ 0\leq p_\tau(\alpha)\leq 1$ for all $\alpha\in{\mathcal U}_\tau$, (ii) $ p_\tau (0_\tau) = 0$, $p_\tau (1_\tau) =1 $ and (iii) it is countably additive in the sense that, given a sequence $\alpha_1,\alpha_2,\ldots$ of pairwise disjoint elements in ${\mathcal U}_\tau$, we have					
\begin{equation}
\label{eq: atlf7}
p_\tau \left(\vee_{i} \alpha_i\right) = \sum_{i} p_\tau (\alpha_i).
\end{equation}			
 An observable at `time' $\tau$ is a map $A_\tau: B(R)\rightarrow{\mathcal U}_\tau$ [where $B(R)$ is the $\sigma$-algebra of Borel subsets of $R$] such that (i) $A_\tau(\phi) = 0_\tau$,  $A_\tau(R) = 1_\tau$, (ii) given disjoint sets $E, F$ in $B(R)$, we have $A_\tau(E)$ and $A_\tau(F)$ disjoint in ${\mathcal U}_\tau$; (iii) if $E_1,E_2,\cdots$ is a sequence of pairwise disjoint sets in $B(R)$, we have
\begin{equation}
\label{eq: atlf8}		
A_\tau\left(\cup_k E_k \right) = \bigvee_k A_\tau(E_k).
\end{equation} 				


We can now state the last two axioms.
\begin{enumerate}
\item[]{\bfseries\emph{$A_4$:(Temporal Evolution)}} : The temporal evolution of a system with history system $({\mathcal U},{\mathcal T},\sigma)$ is given, for each pair $\tau,\tau'\in{\mathcal N}({\mathcal U})$ such that $\tau\lhd\tau'$, by a set of mappings $V(\tau',\tau)$ of ${\mathcal U}_\tau$ onto ${\mathcal U}_{\tau'}$, which are  logic homomorphisms (not necessarily injective) and which satisfy the composition rule $V(\tau'',\tau')\cdot V(\tau',\tau) = V(\tau'',\tau)$ whenever $\tau\lhd\tau', \tau'\lhd\tau''$ and $\tau\lhd\tau''$.

\item[]{\bfseries\emph{$A_5$(a):(Decoherence Functionals)}}: Given a state $p_0\in{\mathcal S}(\tau_0)$ for some $\tau_0\in{\mathcal N}({\mathcal T})$ and a law of evolution $V(\tau',\tau)$, we have a decoherence functional $d = d_{p_0,V}$ which is a mapping from (a subset of) $\Omega\times\Omega$ into $C$ such that
\alphaeqn
\begin{alignat}{2}	
\mbox{(i)}& \hspace{3.5cm}d(\underline{\alpha},\underline{\beta})^{*}  = 
d(\underline{\beta},\underline{\alpha}) & \mbox{  (hermiticity)} \label{eq: atlf9a}\\ 
\mbox{(ii)}& \hspace{3.5cm}d(\underline{\alpha},\underline{\alpha})\geq 0 & \mbox{(positivity)} \label{eq: atlf9b}
\end{alignat}				
\begin{enumerate}
\item[(iii)] If $\underline{\alpha},\underline{\beta}, \cdots$ is an at most countable collection of pairwise disjoint elements of $\Omega$, we have 
\begin{equation}
\label{eq: atlf10}	
\begin{split}					
d(\underline{\alpha}\oplus\underline{\beta}\oplus\cdots ,\underline{\gamma}) = &\quad d(\underline{\alpha}, \underline{\gamma}) + d(\underline{\beta},\underline{\gamma}) + \cdots  \\ & \mbox{(countable additivity)} \\
\end{split} 
\end{equation}				
\end{enumerate}				
\begin{equation}
\label{eq: atlf11}
\mbox{(iv)}\hspace{3.6cm}d(\underline{1},\underline{1}) = 1 \hspace{1cm}\mbox{(normalization)}.
\end{equation}	
\reseteqn
\end{enumerate}
The space of decoherence functionals is denoted as ${\mathcal D}$. In the case of Hilbert space based theories and classical mechanics, explicit expressions for decoherence functional $d_{p_0,V}$ are given in~\cite{dj1}.

A complete set ${\mathcal C}$ of history propositions is called (weakly)
 decoherent (or consistent) with respect to a decoherence functional $d$ if
\begin{equation}
\label{eq: atlf12}		
\mbox{Re}[d(\underline{\alpha},\underline{\beta})] = 0, \hspace{0.4in}\underline{\alpha}, \underline{\beta}\in{\mathcal C}, \hspace{.2in}\underline{\alpha} \neq \underline{\beta}.
\end{equation}	

\begin{enumerate}						
\item[]{\bfseries\emph{$A_5$(b):(Probability Interpretation)}}: The probability that a history $\underline{\alpha}$ in a complete set ${\mathcal C}$ which is decoherent with respect to a decoherence functional $d$ is realized is given by
\begin{equation}
\label{eq: atlf13}
P(\underline{\alpha}) = d(\underline{\alpha},\underline{\alpha}).
\end{equation}
\end{enumerate}				
For $\underline{\alpha},\underline{\beta}\in{\mathcal C}$, we have the classical probability sum rule:
\begin{eqnarray}
\label{eq: atlf14}		
P(\underline{\alpha}\oplus\underline{\beta}) & = & d(\underline{\alpha}\oplus\underline{\beta}, \underline{\alpha}\oplus\underline{\beta})\nonumber \\
 & = & d(\underline{\alpha},\underline{\alpha}) + d( \underline{\beta}, \underline{\beta}) + 2\,\mbox{Re } d(\underline{\alpha},\underline{\beta})\nonumber \\
& = & P(\underline{\alpha}) +P(\underline{\beta}),
\end{eqnarray}
and recalling Eq.(\ref{eq: atlf6})
\begin{eqnarray}
\label{eq: atlf15}			
1 & = d(\underline{1},\underline{1}) & = d(\underline{\alpha}\oplus\underline{\beta}\oplus\cdots, \underline{\alpha}\oplus\underline{\beta}\oplus\cdots)  \nonumber \\
 & & = \sum_{\underline{\alpha}\in{\mathcal C}} P(\underline{\alpha}).
\end{eqnarray}


\section{Symmetries in histories-based generalized quantum mechanics}
\label{sec: sihbgqm}

We shall start with a quick look at the paper by Houtappel {\it et al.}~\cite{hvw1}.


\subsection{Symmetries in the Houtappel {\it et al.} approach}
\label{subsec: siha}

Houtappel {\it et. al} define an invariance transformation as an invertible mapping 
\begin{equation}
\label{eq: siha1}			
\alpha\leftrightarrow\overline{\alpha},\beta\leftrightarrow\overline{\beta},\cdots
\end{equation}				      
(recall that the symbol $\alpha$ implicitly includes the time $t_\alpha$ of the measurement $\alpha$) which leaves the conditional probabilities ($\Pi$-function) invariant:
\begin{equation}
\label{eq: siha2}
\Pi (\overline{\zeta},r_{\overline{\zeta}};\ldots|\ldots;\overline{\nu},r_{\overline{\nu}}) = \Pi \left(\zeta,r_\zeta;\ldots|\ldots;\nu,r_\nu\right).
\end{equation}				 	

Houtappel {\it et al.} explored some consequences of this definition in the classical (Newtonian) mechanics of point particles, relativistic mechanics of point particles, and in quantum theory. From our point of view, the main result in the paper is the generalization of Wigner's theorem given below. In the statement of this theorem, the symbol $\alpha$ of Eq. (\ref{eq: siha1}) represents a decision measurement (Yes-No Experiment) denoted by a pair $(P_\alpha,t_\alpha)$ where $P_\alpha$ is a $1$-dimensional projection operator and $t_\alpha$ is the time of measurement.
		
\noindent\emph{ Generalized Wigner's Theorem}: A mapping of decision measurements onto decision measurements $(P_\alpha,t_\alpha\rightarrow\overline{P}_\alpha,\overline{t}_\alpha)$ leaves the $\Pi$-function invariant if and only if the following two conditions apply:
\begin{enumerate}
\item[(a)] $\overline{P}_\alpha = U P_\alpha U^{-1}$ where $U$ is a unitary or antiunitary operator mapping bijectively coherent subspaces (of the quantum mechanical Hilbert space ${\mathcal H}$ of the system in question) onto coherent subspaces.
\item[(b)] The time order of measurements is either preserved or reversed by the mapping.
\end{enumerate}
We draw two conclusions from the foregoing:
\begin{enumerate}
\item[(i)] Invariance of diagonal elements $d(\alpha,\alpha)$ [see Eqs. (\ref{eq: atlf13}), (\ref{eq: siha2})] of decoherence functionals must be a part of our definition of symmetry (or an implication of it).
\item[(ii)] One should generally expect a two-way connection between Wigner-type symmetries and symmetries defined in the language of histories.
\end{enumerate}

It is of some relevance here to note the distinction between symmetries of the  formalism (unitary/antiunitary transformations in quantum mechanics and canonical transformations in classical mechanics) and symmetries of dynamics (subclass of symmetries of the formalism leaving the Hamiltonian invariant). Not every symmetry of the formalism need be a symmetry of some given Hamiltonian. For example, the parity operator $P$ given by $(Pf)(x) = f(-x)$ is a unitary operator in $L^{2}(R)$; however, $PHP^{-1}\neq H$ for $H=-\frac{\partial^2}{\partial x^2} + \sin x $.

An interesting point to note in the theorem stated above is that the invariance condition does \emph{not} imply a symmetry of dynamics contrary to what one might intuitively expect. (After all, histories or the $\Pi$-functions are supposed to contain all information about dynamics.)

This can also be seen explicitly by having a closer look at Eq. (\ref{eq: intro2}) for $d(\alpha,\alpha)$. A common unitary transformation on all the ingredients -the initial density operator $\rho(t_0)$, the projectors $\alpha_{t_j}$ and the evolution operators $U(t_j,t_k)$ - leaves $d(\alpha,\alpha)$ invariant [in fact it leaves $d(\alpha,\beta)$ of Eq. (\ref{eq: intro5}) invariant]; it does not have to leave  the Hamiltonian invariant to achieve this.


\subsection{Symmetries in the augmented temporal logic formalism}	
\label{subsec: siatlf}

We shall use the following notations:
\begin{enumerate}
\item[] {\sf S} = $({\mathcal U},{\mathcal T},\sigma)$ ({\bfseries\emph{History System}}),
\item[] $\tilde{\mbox{\sf S}}$ = $(\tilde{\mathcal U},\tilde{\mathcal T},\tilde{\sigma})$ ({\bfseries\emph{Standardized History System}}),
\item[] ({\sf S}) = $({\mathcal U},{\mathcal T},\sigma,\Omega,{\mathcal D})$ ({\bfseries\emph{Augmented History System}}).
\end{enumerate}

We define a {\bfseries\emph{morphism}} from {\sf S} into {\sf S'} as a pair $\Phi=(\Phi_1,\Phi_2)$ of mappings such that
\begin{enumerate}
\item[(i)] $\Phi_1:{\mathcal T}\rightarrow{\mathcal T}'$ is a psg homomorphism or antihomomorphism, \emph{i.e.} it satisfies either (a) or (b) below.
\begin{enumerate}
\alphaeqn
\item[(a)] $\tau_1\lhd\tau_2$ implies $\Phi_1(\tau_1)\lhd\Phi_1(\tau_2)$ and 
\begin{equation}
\label{eq: siatlf1}		
\Phi_1(\tau_1\circ\tau_2) = \Phi_1(\tau_1)\circ\Phi_1(\tau_2).
\end{equation}				
\item[(b)] $\tau_1\lhd\tau_2$ implies $\Phi_1(\tau_2)\lhd\Phi_1(\tau_1)$ and
\begin{equation}
\label{eq: siatlf2}			
\Phi_1(\tau_1\circ\tau_2) = \Phi_1(\tau_2)\circ\Phi_1(\tau_1).
\end{equation}				
\reseteqn
\end{enumerate}
\item[(ii)] $\Phi_2:{\mathcal U}\rightarrow{\mathcal U}'$ is a psg homomorphism or antihomomorphism in accordance with (i) (\emph{i.e.} $\Phi_1$ and $\Phi_2$ are either both  homomorphisms or both antihomomorphisms).
\item[(iii)] The following diagram is commutative: 
\begin{equation}
\label{eq: siatlf3}
\begin{CD}
{\mathcal U}@>\Phi_2>>{\mathcal U}' \\
@V{\sigma}VV   @VV{\sigma' \hspace{.4cm} i.e. \hspace{.5cm} \Phi_1\circ
\sigma = \sigma'\circ\Phi_2.}V \\
{\mathcal T}@>>\Phi_1>{\mathcal T}'
\end{CD}
\end{equation}
Here the symbol $\circ$ denotes composition of mappings.
 
Writing $\Phi_1(\tau)=\tau'$ and $\Phi_2|{\mathcal U}_\tau=\Phi_{2\tau}$, the restriction of $\Phi_2$ to the space ${\mathcal U}_\tau$, Eq. (\ref{eq: siatlf3}) implies that $\Phi_{2\tau}$ maps ${\mathcal U}_\tau$ into ${\mathcal U}'_{\tau '}$.
\item[(iv)] Each mapping $\Phi_{2\tau}:{\mathcal U}_\tau\rightarrow{\mathcal U}'_{\tau '}$ is morphism of logics \emph{i.e.} it is injective, preserves partial order, meet, join, and orthocomplementation and maps null and unit elements onto null and unit elements respectively.
\end{enumerate}

A morphism is called an {\bfseries\emph{isomorphism}} if the mappings $\Phi_1$ and $\Phi_2$ are bijective. An isomorphism of {\sf S} onto itself is called an {\bfseries\emph{automorphism}} of {\sf S}. The family of all automorphisms of {\sf S} forms a group called Aut({\sf S}).

The next step is to obtain an induced morphism $\tilde\Phi$=($\tilde\Phi_1$,$\tilde\Phi_2$) of $\tilde{\mbox{\sf S}}$ onto $\tilde{\mbox{\sf S}}'$. This appears to go through smoothly only if $\Phi$ is an isomorphism, which we henceforth assume to be.

Since each $\Phi_{2\tau}$ is a bijection preserving the logic structure, in particular the null and unit elements, $\Phi_2$ maps null history filters onto null history filters and unit history filters onto unit history filters. It follows that $\Phi_2$ induces a bijective mapping $\tilde{\Phi}_2$ of $\tilde{\mbox{\sf S}}$ onto $\tilde{\mbox{\sf S}}'$ which maps $\tilde{0}$ to $\tilde{0}'$, $\tilde{1}$ to $\tilde{1}'$ and typical elements to typical elements; in fact, it is a psg isomorphism.

Similarly, $\Phi_1$ induces a psg isomorphism $\tilde{\Phi}_1$ of $\tilde{\mathcal T}$ onto $\tilde{\mathcal T}'$ and the pair $\tilde{\Phi} = (\tilde{\Phi}_1, \tilde{\Phi}_2)$ gives an isomorphism of $\tilde{\mbox{\sf S}}$ onto $\tilde{\mbox{\sf S}}'$.

The isomorphism $\tilde{\Phi}_2$ preserves the operations $\leq$, $\bot$, $\oplus$ and $\wedge$ defined on $\tilde{\mathcal U}$. The condition of weak disjointness is also preserved. Hence we have an induced mapping $\underline{\Phi}_2$ : $\Omega\rightarrow\Omega'$ mapping $\underline{0}$ to  $\underline{0}'$,  $\underline{1}$ to  $\underline{1}'$ and generic elements to generic elements. 

All the structural properties going into the definition of various operations in $\Omega$ ($\leq$, $\bot$, $\oplus$) are preserved by $\underline{\Phi}_2$. In particular 	
\begin{enumerate}
\item[(a)] $\underline\alpha\bot\underline\beta$\quad if and only if \quad $\underline{\Phi}_2(\underline\alpha)\bot\underline{\Phi}_2(\underline\beta)$
\item[(b)] $\underline{\Phi}_2(\underline\alpha\oplus\underline\beta) = \underline{\Phi}_2(\underline\alpha)\oplus \underline{\Phi}_2(\underline\beta)$
\item[(c)] $\underline{\Phi}_2(\neg\underline\alpha) = \neg \underline{\Phi}_2(\underline\alpha)$.
\end{enumerate}
Henceforth we restrict ourselves to the case ${\mathcal U}'={\mathcal U}$, ${\mathcal T}'={\mathcal T}$; this implies $\tilde{\mathcal U}'=\tilde{\mathcal U}$, $\tilde{\Omega}'=\tilde{\Omega}$, \emph{etc}.

To obtain the transformation law of decoherence functionals we need transformation laws of states and evolution maps. A state $p_\tau\in{\mathcal S}(\tau)$ transforms under $\Phi$ to a state $p'_{\tau '}\in{\mathcal S}(\tau ')$ [where $\tau '=\Phi_1(\tau)$] given by
\begin{equation}
\label{eq: siatlf4}		
p'_{\tau '}(\beta) = p_\tau\left[\Phi^{-1}_{2\tau}(\beta)\right]\quad \mbox{for all}\,\beta\in{\mathcal U}'_{\tau '}.				
\end{equation}			
We can see from Eq. (\ref{eq: siatlf4}) that the mapping $p_\tau\rightarrow p'_{\tau '}$ preserves convex combinations; in particular, it transforms pure states to pure states. An observable $A_\tau\in{\mathcal O}(\tau)$ transforms under $\Phi$ to $A'_{\tau '}\in{\mathcal O}(\tau ')$ given by
\begin{equation}
\label{eq: siatlf5}		
A'_{\tau '}(E) = \Phi_{2\tau}\left[A_\tau(E)\right]\quad \mbox{for all}\,E\in B(R).
\end{equation}		
Given $\tau_1\lhd\tau_2$ in ${\mathcal N}({\mathcal T})$ and an evolution map $V(\tau_2, \tau_1):{\mathcal U}_{\tau_1}\rightarrow{\mathcal U}_{\tau_2}$, $\Phi$ induces an evolution map $V'(\tau'_2 ,\tau_1'):{\mathcal U}_{\tau'_1}\rightarrow{\mathcal U}_{\tau_2 '}$ such that the following diagram is commutative:
\begin{displaymath}			
\begin{CD}
{\mathcal U}_{\tau_1}@>V(\tau_2,\tau_1)>>{\mathcal U}_{\tau_2} \\
@V{\Phi_{2\tau_1}}VV   @VV{\Phi_{2\tau_2}} V \\
{\mathcal U}_{\tau'_1}@>>V'(\tau'_2,\tau'_1)>{\mathcal U}_{\tau'_2}
\end{CD}
\end{displaymath}	
\begin{eqnarray}
\label{eq: siatlf6}
\mbox{\it i.e.  } & V'(\tau'_2,\tau'_1)\circ\Phi_{2\tau_1} = \Phi_{2\tau_2}\circ V(\tau_2,\tau_1). &	
\end{eqnarray}				
The pair $\Phi=(\Phi_1,\Phi_2)$ induces a map $\Phi_3:{\mathcal D}\rightarrow{\mathcal D}$, given by
\begin{equation}
\label{eq: siatlf7}		
\Phi_3 \left(d_{p_0,V}\right) = d_{p'_{0},V'}
\end{equation}				
where $	p'_{0}$, $V'$ are given by Eqs. (\ref{eq: siatlf4}) and (\ref{eq: siatlf6}). Eq. (\ref{eq: siatlf7}) can be used in a formal treatment of symmetries whether or not an explicit expression for the decoherence functionals is available.

We define a \emph{symmetry operation} for the augmented history system ({\sf S}) (see notation above) as a triple $\underline\Phi=(\Phi_1,\Phi_2,\Phi_3)$ such that
\begin{itemize}							
\item[(i)] $\Phi=(\Phi_1,\Phi_2)$ is an automorphism of {\sf S};
\item[(ii)] $\Phi_3:{\mathcal D}\rightarrow{\mathcal D}$ satisfies the condition
\begin{equation}
\label{eq: siatlf8}		
\mbox{Re}\left[\Phi_3(d)\left(\underline\Phi_2(\underline\alpha),\underline\Phi_2(\underline\beta)\right)\right] = \mbox{Re}[d(\underline\alpha,\underline\beta)] \hspace{1cm} \mbox{for all } \underline\alpha,\underline\beta\in\Omega.    
\end{equation}		
\end{itemize}   
Note that the condition (\ref{eq: siatlf8}) implies preservation of the (weak) decoherence condition (\ref{eq: atlf12}) as well as the probability expression (\ref{eq: atlf13}).\footnote{A stronger decoherence condition often employed is $d(\alpha,\beta) = 0$ for $\alpha\neq\beta$~\cite{gh1,jh1,jh2}. The appropriate stronger invariance condition replacing Eq. (\ref{eq: siatlf8}) would then be $\Phi_3 (d) \left(\underline{\Phi}_2 (\underline{\alpha}),\underline{\Phi}_2(\underline{\beta})\right) = d\left(\underline{\alpha},\underline{\beta}\right).$}  A symmetry operation will be called \emph{orthochronous} if the mappings $(\Phi_1,\Phi_2)$ are isomorphisms and \emph{nonorthochronous} if they are anti-isomorphisms. The former preserves the `temporal order' of `events' while the latter reverses it.	
			
A symmetry operation as defined above is a symmetry of the formalism. A symmetry of dynamics would be a member of the subclass of these symmetries leaving the evolution map $V(.,.)$ invariant, \emph{i.e.} satisfying the condition
\begin{equation}
\label{eq: siatlf9}		
V'(\tau_2,\tau_1) = V(\tau_2,\tau_1),
\end{equation}				
for all $\tau_2,\tau_1$ such that $\tau_1\lhd\tau_2$. This equation can be meaningful only if the two sides define mappings between the same spaces; this implies ${\mathcal U}_{\tau_2'}\approx{\mathcal U}_{\tau_2}$ and  ${\mathcal U}_{\tau_1'}\approx{\mathcal U}_{\tau_1}$ where $\approx$ indicates isomorphism (of logics). Since $\tau_1$ and $\tau_2$ are fairly arbitrary (subject only to the condition $\tau_1\lhd\tau_2$), it appears that, in the class of theories being discussed, symmetries of dynamics are definable only for the subclass in which all ${\mathcal U}_\tau$'s are isomorphic.

\noindent Remark: Given a pair of spaces $({\mathcal U},{\mathcal T})$, there may be more than one possible candidates for the decoherence functionals $d_{p,V}$. The family of transformations constituting symmetries will then be correspondingly different for different $d_{p,V}$'s. This is due to the invariance condition (\ref{eq: siatlf8}) which varies with the choice of  $d_{p,V}$.


\subsection{Symmetries in general quasitemporal theories and in Isham-Linden type theories}
\label{subsec: sigqt}

In going from the augmented temporal logic formalism to general quasitemporal theories and from there to the Isham-Linden type theories, one has to drop some structures along the way. In the first transition, we have to drop the logic structure of ${\mathcal U}_\tau$'s and, in the second, the quasitemporal structure itself. The definition of symmetry in these theories must ensure preservation of the remaining mathematical structure.

A concrete quasitemporal theory must define an embedding of ${\mathcal U}$ in $\Omega$ in explicit terms. Such a theory has associated with it an augmented history system, ({\sf S})= $({\mathcal U},{\mathcal T},\sigma,\Omega,{\mathcal D})$. Here, as noted in the previous section, the triple $({\mathcal U},{\mathcal T},\sigma)$ is the analogue of $(\tilde{\mathcal U},\tilde{\mathcal T},\tilde\sigma)$ above. A symmetry operation in such a theory may be defined as a triple $\Phi=(\Phi_1,\Phi_2,\Phi_3)$ of (invertible) mappings satisfying the conditions stated above except that
\begin{itemize}								
\item[(i)] the condition (iv) on $\Phi_{2\tau}$ must be replaced by an appropriate structure preserving mapping if a definite mathematical structure (of an algebra, for example) is identified for the spaces of single `time' histories in the formalism; otherwise it should be dropped.
\item[(ii)] the mapping $\Phi_3$ now cannot be specified as in Eq. (\ref{eq: siatlf7}) and must be treated as an independent mapping.
\end{itemize}

In the Isham-Linden type theories, a symmetry may be defined as a pair $(\underline\Phi_2,\Phi_3)$ of invertible mappings, $\underline\Phi_2:\Omega\rightarrow\Omega$ and $\Phi_3:{\mathcal D}\rightarrow{\mathcal D}$, having the properties described earlier (recall, in particular, that $\underline\Phi_2$ preserves the mathematical operations in $\Omega$) and satisfying the invariance condition (\ref{eq: siatlf8}) (or the stronger version of it).

If, in any of the types of theories discussed above, a concrete expression for the decoherence functional is available, the mappings involved in the definition of symmetry can generally be shown to belong to well defined classes. This was the situation in the work of Houtappel {\it et al.} described above (where we had the $\Pi$-functions instead of the decoherence functional) and prevails in sections~\ref{sec: sinqm} and~\ref{sec: sihvocm} and in the work of Schreckenberg~\cite{ss1} briefly described below.

For Isham-Linden type theories in which $\Omega$ is the lattice ${\mathcal P}({\mathcal V})$ of projection operators in a finite dimensional Hilbert space ${\mathcal V}$ (of dimension $>$ 2), Isham {\it et al.}~\cite{il1} proved that every decoherence functional $d: \Omega\times\Omega\rightarrow C$ can be written as
\begin{equation}
\label{eq: sigqt1}		
d(\underline\alpha,\underline\beta) = \mbox{tr}_{{\mathcal V}\otimes{\mathcal V}}\left[(\underline\alpha\otimes\underline\beta)X\right].
\end{equation}				
Here $X$ belongs to a class ${\mathcal X}_{\mathcal D}$ of operators in the space ${\mathcal V}\otimes{\mathcal V}$ satisfying a definite set of conditions of hermiticity, positivity and normalization chosen so that $d$ of Eq. (\ref{eq: sigqt1}) satisfies the usual conditions. 
In~\cite{ss1}, `physical symmetries' were defined in the framework of theories of the above sort as affine one-to-one maps from $[{\mathcal P}({\mathcal V})\otimes{\mathcal P}({\mathcal V})]\times{\mathcal X}_{\mathcal D}$ into itself preserving the quantity on the right hand side of Eq. (\ref{eq: sigqt1}). A Wigner type theorem~\cite{blot1,ss2} was proved there showing that the `physical symmetries' are in one-to-one correspondence with the so-called `homogeneous symmetries' (those implemented by the operators of the form $U\otimes U$ on ${\mathcal V}\otimes{\mathcal V}$ where $U$ is a unitary or antiunitary operator on ${\mathcal V}$). Here a symmetry operation can be easily seen to be described as a pair $(\underline\Phi_2,\Phi_3)$ where the mappings $\underline\Phi_2:{\mathcal P}({\mathcal V})\rightarrow{\mathcal P}({\mathcal V})$ and $\Phi_3:{\mathcal X}_{\mathcal D}\rightarrow{\mathcal X}_{\mathcal D}$ are those in Eqs. (II.17),(II.18) and (III.7) of~\cite{ss1}. In~\cite{ss2} symmetries of individual decoherence functionals [maps $\underline\alpha\rightarrow\underline\alpha'=U\underline\alpha U^{\dagger}$, $\underline\beta\rightarrow\underline\beta'$ satisfying the condition $d(\underline\alpha',\underline\beta')=d(\underline\alpha,\underline\beta)$ for all $\underline\alpha, \underline\beta$ in $\Omega ={\mathcal P}({\mathcal V})$ for a given $d$] were considered in some detail.


\section{Traditional vs. temporal logic descriptions of symmetries in nonrelativistic quantum mechanics}
\label{sec: sinqm}

In this section we shall establish a reciprocal relationship between the traditional description of symmetry operations in nonrelativistic quantum mechanics and those in the present formalism. The transition from traditional Wigner symmetries to those in the formalism of the previous section is described most transparently in the HPO (History Projection Operator) formalism~\cite{cji1}.


\subsection{The HPO formalism}
\label{subsec: hpof}

If histories are represented as (multi-time) propositions, it is natural to look for a representation of histories as projection operators in some Hilbert space. This is achieved for traditional quantum theory in~\cite{cji1}. In this case, we first construct the Cartesian product
\begin{equation}
\label{eq: hpof1}
{\mathcal V}=\prod\limits_{t\in R}{\mathcal H}_t
\end{equation}		
of (naturally isomorphic) copies ${\mathcal H}_t$ of the quantum mechanical Hilbert space ${\mathcal H}$ of the system. Let $w=(w_t)$ be a fixed vector in ${\mathcal V}$ such that $\parallel w_t\parallel=1$ for all $t\in R$. Let ${\mathcal F}$ be the subspace of ${\mathcal V}$ consisting of vectors $v$ such that $v_t = w_t$ for all but a finite set of $t$-values. The scalar product $(.,.)$ on ${\mathcal H}$ induces the following scalar product on ${\mathcal F}$:
\begin{equation}
\label{eq: hpof2}		
(v',v)_{\mathcal F} = \prod\limits_{t\in R} (v'_t,v_t)_{{\mathcal H}_t}.
\end{equation}				
The completion $\tilde{\mathcal H}$ of this inner product space is the desired Hilbert space; it is the infinite tensor product
\begin{equation}
\label{eq: hpof3}		
\tilde{\mathcal H} = \otimes^{w}_{t\in R} {\mathcal H}_t.
\end{equation}				
The history $\alpha$ of Eq. (\ref{eq: intro1}) is represented in $\tilde{\mathcal H}$ by the homogeneous projection operator 
\begin{equation}
\label{eq: hpof4}		
\overline{\alpha} = \alpha_{t_1}\otimes\alpha_{t_2}\otimes\cdots\otimes\alpha_{t_n}.
\end{equation} 				
Given another projection operator $\overline{\beta}$ in $\tilde{\mathcal H}$ representing some history $\beta$ such that $\overline{\alpha}$ and  $\overline{\beta}$ are mutually orthogonal, the inhomogeneous projection operator $\overline{\alpha}+\overline{\beta}$ can be taken to represent the history proposition `$\alpha$ or $\beta$'. Histories are referred to as `homogeneous' or `inhomogeneous' depending on whether they are represented by homogeneous or inhomogeneous projection operators. Inclusion of inhomogeneous histories facilitates the introduction of the logical operation of negation of a history proposition. Given $\alpha$ and $\overline{\alpha}$ as above, the projection operator representing the negation of the history proposition $\alpha$ is $\tilde{I}-\overline{\alpha}$ where $\tilde{I}$ is the identity operator on $\tilde{\mathcal H}$.  


\subsection{Symmetries in traditional quantum mechanics in the HPO formalism}
\label{subsec: sitqmihf}

We denote by ${\mathcal P}_1({\mathcal H})$ the space of one-dimensional projection operators (\emph{i.e.} operators of the form $P_\Psi = |\Psi\rangle\langle\Psi|$ on the quantum mechanical Hilbert space ${\mathcal H}$). In terms of these operators the transition probability formula reads 
\begin{equation}
\label{eq: sitqmihf1}		
P (|\Psi\rangle \rightarrow|\Phi\rangle)=|\langle\Phi|\Psi\rangle|^2=\mbox{Tr}(P_\Psi P_\Phi).
\end{equation}				
According to Wigner's theorem~\cite{blot1} given a bijection ${\mathcal P}_1({\mathcal H})\rightarrow{\mathcal P}_1({\mathcal H})$ ($P\rightarrow P'$) which preserves transition probabilities, there exists a unitary or antiunitary operator $U$ on ${\mathcal H}$ such that 
\begin{equation}
\label{eq: sitqmihf3}		
P'= U P U^{-1} \hspace{2cm}\mbox{for all}\hspace{.2cm}P\in{\mathcal P}_1({\mathcal H}).
\end{equation}				
Recall that, the action of such a $U$ on ${\mathcal B}({\mathcal H})$ (the algebra of bounded operators on ${\mathcal H}$) is given by 
\alphaeqn
\begin{eqnarray}	
A \rightarrow A' =&  U A U^{-1}  &\mbox{ for unitary}\quad U  \label{eq: sitqmihf4a}\\
A \rightarrow A' =&  U A^{\dagger} U^{-1} &\mbox{ for antiunitary}\quad U. \label{eq: sitqmihf4b}
\end{eqnarray}				
\reseteqn
If $A$ is self-adjoint, the expressions on the right side in Eqs. (\ref{eq: sitqmihf4a}), (\ref{eq: sitqmihf4b}) are the same; in particular, this is the case for a projection operator and a density operator.

A unitary/antiunitary operator $U$ on ${\mathcal H}$ defines a unitary/antiunitary operator $\tilde{U}$ on $\tilde{\mathcal H}$ such that 
\begin{equation}
\label{eq: sitqmihf5}		
\tilde{U}\left[\Psi_{t_1}\otimes\Psi_{t_2}\otimes\cdots\otimes\Psi_{t_n}\right] = U\Psi_{t_1}\otimes U\Psi_{t_2}\otimes\cdots\otimes U \Psi_{t_n}.
\end{equation}				
The homogeneous projection operator $\overline{\alpha}$ of Eq. (\ref{eq: hpof4}) transforms into
\begin{equation}
\label{eq: sitqmihf6}		
\overline{\alpha}' = \tilde{U}\overline{\alpha}\tilde{U}^{-1} = \alpha'_{t_1}\otimes\cdots\otimes\alpha'_{t_n}
\end{equation}				
where
\begin{equation}
\label{eq: sitqmihf7}	 	
\alpha'_{t_j} = U \alpha_{t_j} U^{-1}.
\end{equation}				
Given $\overline{\alpha}$ of Eq. (\ref{eq: hpof4}) and $\overline{\beta}=\beta_{s_1}\otimes\cdots\otimes\beta_{s_m}$ such that $t_n < s_1$, the composite history projector $\overline{\alpha}\circ\overline{\beta} = \alpha_{t_1}\otimes\cdots\otimes\alpha_{t_n}\otimes\beta_{s_1}\otimes\cdots\otimes\beta_{s_m}$, being a homogeneous projector, transforms as in Eq. (\ref{eq: sitqmihf6}), giving
\begin{equation}
\label{eq: sitqmihf9}		
\overline{\alpha}\circ\overline{\beta} \rightarrow \tilde{U}(\overline{\alpha}\circ\overline{\beta})\tilde{U}^{-1} = (\tilde{U}\overline{\alpha}\tilde{U}^{-1})\circ(\tilde{U}\overline{\beta}\tilde{U}^{-1}).
\end{equation}				
The transformation law (\ref{eq: sitqmihf6}), therefore, defines an automorphism of the psg ${\mathcal K}_2$ (the space of history filters ${\cal U}$ in the present case). This action trivially extends to inhomogeneous projectors:
\begin{equation}
\label{eq: sitqmihf8}		
\overline{\alpha}+\overline{\beta}\rightarrow\overline{\alpha}'+\overline{\beta}' = \tilde{U}(\overline{\alpha}+\overline{\beta})\tilde{U}^{-1}.
\end{equation}				

In writing Eq. (\ref{eq: sitqmihf9}), we have implicitly assumed that there is no transformation of time involved. There may, in general, be a transformation of the time variable also involved,
\begin{equation}
\label{eq: sitqmihf10}		
t\rightarrow t' = f(t).		
\end{equation}				
If $t'$ is to serve as a time variable, the function $f$ must be monotone. There are two possibilities:
\begin{enumerate}
\item[(i)] $f$ is monotone increasing. In this case $t_i < t_j$ implies $t'_i < t'_j$. Given temporal supports $A = (t_1,\ldots,t_n)$ and $B = (s_1,\ldots,s_m)$ such that $A\lhd B$ (corresponding to $t_n < s_1$) we have $A'=(t'_1,\ldots,t'_n)$, $B'=(s'_1,\ldots,s'_m)$ and
\begin{equation}
\label{eq: sitqmihf12}		
A' \lhd B' , \hspace{2cm}(A\circ B)' = A'\circ B'
\end{equation}				
giving an automorphism of the psg ${\mathcal K}_1$ (the space of temporal supports ${\mathcal T}$ in the present case).
\item[(ii)] $f$ is monotone decreasing (time reversal $t'=-t$ is an important special case of this). In this case $t_i<t_j$ implies $t'_j<t'_i$. When $A\lhd B$, we have 
\begin{eqnarray}
\label{eq: sitqmihf13}		
B' \lhd  A',  & \hspace{2cm} & (A\circ B)'  =  B' \circ A'
\end{eqnarray}				
giving an antiautomorphism of ${\mathcal K}_1$. In this case Eq. (\ref{eq: sitqmihf13}) implies that Eq. (\ref{eq: sitqmihf9}) must be replaced by an antiautomorphism of ${\mathcal K}_2={\mathcal U}$:
\begin{equation}
\label{eq: sitqmihf15}		
\overline{\alpha}\circ\overline{\beta}\rightarrow\tilde{U}(\overline{\alpha}\circ\overline{\beta})\tilde{U}^{-1} = (\tilde{U}\overline{\beta}\tilde{U}^{-1})\circ(\tilde{U}\overline{\alpha}\tilde{U}^{-1}).
\end{equation}			
\end{enumerate}
Summarizing, we have shown that a Wigner symmetry implemented as in Eq. (\ref{eq: sitqmihf3}) along with the transformation (\ref{eq: sitqmihf10}) of time implies, for the history system $({\mathcal U},{\mathcal T},\sigma)=({\mathcal K}_2,{\mathcal K}_1,\sigma)$ the following:
\begin{enumerate}
\item[(i)] An automorphism or antiautomorphism $\Phi_1$ of ${\mathcal T}$ [given by $\Phi_1(A) = A'$];
\item[(ii)] An automorphism or antiautomorphism $\Phi_2$ of ${\mathcal U}$ [in accordance with (i)] given by
\begin{equation}
\label{eq: sitqmihf16}
\Phi_2(\overline{\alpha}) = \tilde{U}\overline{\alpha}\tilde{U}^{-1}.
\end{equation}			
\item[(iii)] Projections (from ${\mathcal U}$ to ${\mathcal T}$) implied by $\sigma$ are preserved by the mappings $\Phi_1$ and $\Phi_2$ making the following diagram commutative:
\begin{equation}
\label{eq: sitqmihf17}
\begin{CD}
{\mathcal U}@>\Phi_2>>{\mathcal U} \\
@V{\sigma}VV  @VV{\sigma \hspace{.3cm} i.e. \hspace{.4cm} \Phi_1\circ\sigma
=\sigma\circ\Phi_2.}V \\
{\mathcal T}@>>\Phi_1>{\mathcal T}	
\end{CD}
\end{equation}				
Eq. (\ref{eq: sitqmihf17}) implies that, for any $t\in R$ [=${\mathcal N}({\mathcal T})$ in the present case], $\Phi_2$ maps ${\mathcal U}_t={\mathcal P}({\mathcal H}_t)$ into ${\mathcal U}_{t'}={\mathcal P}({\mathcal H}_{t'})$ where $t'=\Phi_1(t)=f(t)$.
\item[(iv)] For each $t\in R ={\mathcal N}({\mathcal T})$, the mapping $\Phi_{2t}=\Phi_2\mid{\mathcal U}_t$ is an isomorphism of the logic ${\mathcal U}_t ={\mathcal P}({\mathcal H}_t)$ onto ${\mathcal U}_{t'}={\mathcal P}({\mathcal H}_{t'})$.
\item[(v)] Eq. (\ref{eq: sitqmihf8}) and (iv) above imply that the mapping $\Phi_2$ on ${\mathcal U}={\mathcal K}_2$ extends to a bijective mapping $\underline{\Phi}_2$ on the space $\Omega$ of inhomogeneous projectors onto itself preserving the lattice operations in it.
\end{enumerate}				
					
\noindent Note: Whether $\Phi_1$ and $\Phi_2$ are automorphisms or antiautomorphisms depends only on whether $f(t)$ in Eq. (\ref{eq: sitqmihf10}) is monotone increasing or decreasing and not on whether $U$ (and corresponding $\tilde{U}$) is unitary or antiunitary. We can very well have a situation where, for example, $U$ is antiunitary, $\Phi_1$ is the identity mapping and $\Phi_2$ is an automorphism.

The transformation law of the decoherence functionals $d(\alpha,\beta)$ of Eq. (\ref{eq: intro5}) following from that of the initial state $\rho(t_0)$ and the evolution operator $U(t',t)$ gives $\rho(t_0)\rightarrow U\rho(t_0) U^{-1}$ and
\alphaeqn
\begin{equation}
\label{eq: sitqmihf18}		
C_\alpha \rightarrow 
\begin{cases}
U C_\alpha U^{-1}& \text{for $U$ unitary}, \\
U C^{\dagger}_\alpha U^{-1}& \text{for $U$ antiunitary}.
\end{cases}
\end{equation}				
which implies that 
\begin{equation}
\label{eq: sitqmihf19}		
d(\alpha,\beta)\rightarrow
\begin{cases}
d(\alpha,\beta)& \text{for $U$ unitary}, \\
d(\alpha,\beta)^{*}&\text{for $U$ antiunitary}.
\end{cases}
\end{equation}				
\reseteqn
Equation (\ref{eq: sitqmihf19}) describes the mapping $\Phi_3$ of Eq. (\ref{eq: siatlf7}) and is clearly consistent with Eq. (\ref{eq: siatlf8}). 

We have seen above that a traditional Wigner symmetry [supplemented with the transformation (\ref{eq: sitqmihf10}) of time] implies, in the history version, a symmetry as defined in section~\ref{sec: sihbgqm}. We now consider the reverse connection.


\subsection{Recovering Wigner symmetries from those in the temporal logic formalism}
\label{subsec: rws}

The simplest way to obtain Wigner symmetries from those of section~\ref{sec: sihbgqm} is to note that since, for each $t\in R ={\mathcal N}({\mathcal T})$, ${\mathcal U}_t\approx{\mathcal P}({\mathcal H})$, a symmetry of section~\ref{sec: sihbgqm} implies an automorphism of ${\mathcal P}({\mathcal H})$ which, in turn, implies the existence of a unitary/antiunitary mapping on ${\mathcal H}$. (See Theorem 4.27 in~\cite{vsv1}).

Note that the argument given above is independent of the nature of `time'; it goes through if ${\mathcal K}_1$ is replaced by a general space ${\mathcal T}$ of temporal supports consistent with axioms $A_1$ and $A_2$ of section~\ref{sec: hbgqm}. This fact will be used in section~\ref{sec: cbsacl}.

A pedagogically simpler route to recover Wigner symmetries from those of section~\ref{sec: sihbgqm} is to obtain from the latter, the condition of invariance of transition probabilities. To this end (noting that ${\mathcal U}$, ${\mathcal T}$ and $\Omega$ here are the same as in the previous subsection) we apply Eq. (\ref{eq: siatlf8}) to the $\underline{\beta}=\underline{\alpha}={\alpha}$ where $\alpha=\alpha_{t_1}$ is a single-time history (\emph{i.e.} a projection operator on ${\mathcal H}$). From Eqs. (\ref{eq: intro3}) and (\ref{eq: intro5}) we have $C_\alpha = \alpha_{t_1} U(t_1,t_0)$ and
\begin{eqnarray}
\label{eq: rws1}		
d(\alpha,\alpha) & = & \mbox{Tr}\left[C_\alpha\rho(t_0) C^{\dagger}_\alpha\right] \nonumber \\
 & = & \mbox{Tr}\left[\alpha_{t_1} U(t_1,t_0) \rho(t_0) U(t_1,t_0)^{-1}\right].
\end{eqnarray}				
Writing $\Phi_1(t_j)=t'_j$ ($j=0,1$), and $\Phi_2(\alpha)=\alpha' = \alpha'_{t'_j}$, the condition $d'(\alpha',\alpha')=d(\alpha,\alpha)$ gives
\begin{equation}
\label{eq: rws2}		
\mbox{Tr}\left[\alpha'_{t'_1}U'(t'_1,t'_0)\rho'(t'_0)U'(t'_1,t'_0)^{-1}\right] = \mbox{Tr}\left[\alpha_{t_1} U(t_1,t_0) \rho(t_0) U(t_1,t_0)^{-1}\right].
\end{equation}				
Putting $t_1=t_0$ in Eq. (\ref{eq: rws2}), we get		
\begin{equation}
\label{eq: rws3}		
\mbox{Tr}\left[\alpha'_{t'_0}\rho'(t'_0)\right]=\mbox{Tr}\left[\alpha_{t_0}\rho(t_0)\right].
\end{equation}				
Taking $\rho(t_0)$ to be a pure state (one dimensional projection operator), $\rho'(t'_0)$ must also be a pure state. [See the statement after Eq. (\ref{eq: siatlf4}).] Similarly, taking $\alpha_{t_0}$ to be one dimensional projector, $\alpha'_{t'_0}$ must also be a one dimensional projector because the mapping $\Phi_{2t_0}$ preserves the logic structure of ${\mathcal U}_{t_0}={\mathcal P}({\mathcal H})$. With these choices, Eq. (\ref{eq: rws3}) implies invariance of transition probabilities. It follows that we now have a Wigner symmetry.


\section{Traditional vs. temporal logic descriptions of symmetries in classical mechanics}
\label{sec: sihvocm}

For classical mechanics of a system with phase space $\mit\Gamma$, the space of temporal supports is taken, for simplicity, to be ${\mathcal T}={\mathcal K}_1$ [which implies ${\mathcal N}({\mathcal T})=R$, the traditional space for the flow of time]. For each $t\in R$, the space ${\mathcal U}_t$ is an isomorphic copy of $B(\mit\Gamma)$, the Boolean logic of Borel subsets of $\mit\Gamma$. A typical history filter  is of the form (considering, for simplicity, history filters with finite temporal supports only)
\alphaeqn
\begin{equation}
\label{eq: sihvocm1}		
\alpha = \{\alpha_{t_1},\alpha_{t_2},\ldots,\alpha_{t_n}; t_1 < t_2<\ldots< t_n, \alpha_{t_i}\in{\mathcal U}_{t_i}\}
\end{equation}				
which is represented as
\begin{equation}
\label{eq: sihvocm2}		
\alpha = \alpha_{t_1}\circ\alpha_{t_2}\circ\cdots\circ\alpha_{t_n}.
\end{equation}				
\reseteqn
Construction of $\tilde{\mathcal U}$ and $\Omega$ as well as descriptions of operations/relations in them are straightforward.

In this case, states at any time $t$ are the probability measures on the measurable space $\left(\mit\Gamma,B(\mit\Gamma)\right)$. In most concrete situations they are measures associated with density functions in phase space. Observables at any time $t$ are maps $A_t : B(R)\rightarrow B(\mit\Gamma)$ with the properties as stated in section~\ref{sec: hbgqm}. Temporal evolution is given by measurable maps $V(t',t):{\mathcal U}_t\rightarrow{\mathcal U}_{t'}$ which we assume to be bijective. Since single points of $\mit\Gamma$ are elements of $B(\Gamma)$, this defines a bijective map of $\mit\Gamma$ onto itself which we also denote as $V(t',t)$. Given $t_0<t_1<t_2<\ldots<t_n$ and $\xi_{t_0}\in\mit\Gamma$, let $\xi(t_j;\xi_{t_0})$ be the point of $\mit\Gamma$ given by
\begin{equation}
\label{eq: sihvocm3}		
\xi(t_j;\xi_{t_0}) = \left[ V(t_j,t_{j-1})\cdot V(t_{j-1},t_{j-2})\cdot\cdots\cdot V(t_2,t_1)\cdot V(t_1,t_0)\right]\left(\xi_{t_0}\right).		
\end{equation}				
Given a history $\alpha$ as above and another history $\beta=\beta_{t_1}\circ\beta_{t_2}\circ\cdots\circ\beta_{t_n}$, we define the decoherence functional corresponding to the state $p_{t_0}$ and the law of evolution $V(.,.)$ as
\begin{equation}
\label{eq: sihvocm4}		
d_{p_0, V}(\alpha,\beta) = p_{t_0}(E_{\alpha\beta}) =\int_{\mit\Gamma} dp_{t_0}(\xi_{t_0}) K^{\alpha\beta}(\xi_{t_0})
\end{equation}			
where $E_{\alpha\beta}$ is the subset of $\mit\Gamma$ consisting of those points $\xi_{t_0}\in\mit\Gamma$ for which $\xi(t_j;\xi_{t_0})$ lies in $\alpha_{t_j}\cap\beta_{t_j}$, for each $j = 1,2,\ldots ,n$ and $K^{\alpha\beta}$ is the characteristic function of $E_{\alpha\beta}$. Equation (\ref{eq: sihvocm4}) serves  to define the decoherence functional $d(=d_{p_0,V})$ for any pair of finite histories. (The general case of two finite histories with different temporal supports can be reduced to the simpler case of common temporal supports by taking some of the $\alpha_{t_i}$ and/or $\beta_{t_j}$ equal to $\mit\Gamma$.)	

In the traditional formalism of Hamiltonian dynamics (in the setting of symplectic manifolds), symmetries of the formalism are the canonical transformations (diffeomorphisms of $\Gamma$ preserving the symplectic form). The histories-based formalism (at the level of generality kept in this paper), however,does not involve smooth structures and infinitesimal versions of dynamics (Hamilton's equations). It operates in the more general framework of (topological) measure spaces employing Borel sets and Borel measurable evolution maps. The symmetries of the formalism in the present context, therefore, are bijective Borel measurable maps of $\mit\Gamma$ onto itself. 

After these preliminaries, we now consider the two-way connection between symmetries as mentioned above.
\begin{enumerate}
\item[(1)] Given an invertible transformation of the time variable [Eq. (\ref{eq: sitqmihf10})] and a bijective Borel measurable mapping $F:\mit\Gamma\rightarrow\mit\Gamma$ $[\xi\rightarrow\xi' = F(\xi)]$ we construct a symmetry operation $\underline{\Phi}=(\Phi_1,\Phi_2,\Phi_3)$ as follows:
\begin{enumerate}
\item[(i)] An automorphism/antiautomorphism $\Phi_1$ of ${\mathcal K}_1$ is constructed as in section~\ref{sec: sinqm}.
\item[(ii)] The mapping $\Phi_{2t} : B(\mit\Gamma)\rightarrow B(\mit\Gamma)$
 defined by
\begin{equation}
\label{eq: sihvocm5}	
\Phi_{2t}(A) = F(A) 	\hspace{2cm} \mbox{for all } A\in B(\mit\Gamma)
\end{equation}				
is an automorphism of the logic $B(\mit\Gamma)$. Note that the mapping $\Phi_{2t}$ is the same for all $t$.
\item[(iii)] Given $\alpha = (\alpha_{t_1},\alpha_{t_2},\ldots,\alpha_{t_n})\in{\mathcal U}$, we have
\begin{equation}
\label{eq: sihvocm6}		
\Phi_2(\alpha) = \left(\Phi_{2t_1}(\alpha_{t_1}),\ldots,\Phi_{2t_n}(\alpha_{t_n})\right).
\end{equation}
\item[(iv)] The verification that the pair $\Phi=(\Phi_1,\Phi_2)$ is an automorphism of the history system {\sf S} = $({\mathcal U},{\mathcal T},\sigma)$ is straightforward.
\item[(v)] The transformation laws of the states $p_{t_0}$ and the evolution maps $V_{t,s}=V(t,s)$ appearing in the classical decoherence functional Eq. (\ref{eq: sihvocm4}) are obtained using Eqs. (\ref{eq: siatlf3}), (\ref{eq: siatlf4}) and (\ref{eq: sihvocm5}). This gives
\begin{equation}
\label{eq: sihvocm7}		
p'_{t'_0}(\beta) = p_{t_0}\left[\Phi_{2t_0}^{-1}(\beta)\right] = p_{t_0}\left[F^{-1}(\beta)\right]
\end{equation}				
and
\begin{equation}
\label{eq: sihvocm8}		
V'_{t'_2,t'_1}(\xi) = F\left[V_{t_2,t_1}\left(F^{-1}(\xi)\right)\right].
\end{equation}				
The transformed decoherence functional is given by 
\begin{equation}
\label{eq: sihvocm9}		
d'(\alpha',\beta') = p'_{t_0'}(E'_{\alpha'\beta'}) = p_{t_0}[ F^{-1}(E'_{\alpha'\beta'})]
\end{equation}
where
\begin{eqnarray}
\label{eq: sihvocm10}		
E'_{\alpha'\beta'} & = &\{ \xi'_0 \in\mit\Gamma; \xi'(t'_j;\xi'_0)\in\alpha'_{t'_j}\cap\beta'_{t'_j}, j=1,\ldots,n\} \nonumber \\
& = &F\{ \xi_0 \in\mit\Gamma; \xi(t_j;\xi_0)\in\alpha_{t_j}\cap\beta_{t_j}, j=1,\ldots,n\} \nonumber \\
& = & F(E_{\alpha\beta}).
\end{eqnarray}				
Equation (\ref{eq: sihvocm9}) now gives
\begin{equation}
\label{eq: sihvocm11} 		
d'(\alpha',\beta') = p_{t_0}(E_{\alpha\beta}) = d(\alpha,\beta).
\end{equation}
which verifies Eq. (\ref{eq: siatlf8}) in the present case.
\end{enumerate}
\item[(2)] Given a symmetry operation $\underline{\Phi}=(\Phi_1,\Phi_2,\Phi_3)$ we recover the maps $f$ and $F$ as follows:
\begin{enumerate}
\item[(i)] The restriction of $\Phi_1$ to ${\mathcal N}({\mathcal T})=R$ fixes the map $f$.
\item[(ii)] $\Phi_2$ determines, for each $t\in{\mathcal N}({\mathcal T})=R$, $\Phi_{2t} : B(\mit\Gamma)\rightarrow B(\mit\Gamma)$ which, when restricted to single point sets, determines a mapping $F : \mit\Gamma\rightarrow\mit\Gamma$ which is bijective and Borel measurable.
\end{enumerate}
\end{enumerate}


\section{Physical equivalence of histories}				
\label{sec: peoh}

Gell-Mann and Hartle~\cite{gh3} have emphasized the need to understand the nature of physical equivalence between sets of (coarse-grained) histories of a closed system as a prerequisite for a clear understanding of some fundamental questions like what would it mean for the universe to exhibit essentially inequivalent quasi classical realms. Assuming that the spacetime has a fixed geometry permitting foliation in space-like hypersurfaces and that the underlying dynamics of the  universe is governed by a canonical quantum field theory, they considered, in histories-based version of this dynamics, a few notions of physical equivalence of sets of histories. We shall now show that all these notions reduce to special cases of a simple criterion for physical equivalence which can be stated concisely in terms of the symmetry operations described above.

The obvious guiding principle for such a criterion is that histories related through transformations leaving all observable quantities invariant must be treated as physically equivalent. The observable quantities for histories (of a closed system) are the probabilities of (decoherent) histories and (in quasitemporal theories) `temporal order' of `events'. So we propose the following criterion of physical equivalence: All histories related to each other through orthochronous symmetry operations are physically equivalent.		
				
We now take up various notions of physical equivalence considered by Gell-Mann and Hartle. Since, in our description of symmetry $\Phi_3$ is determined in terms of $\Phi_1$ and $\Phi_2$ [see Eq. (\ref{eq: siatlf7})] , it is adequate to give  $\Phi_1$ and $\Phi_2$ corresponding to the various notions of physical equivalence. Extension of $\Phi_2$ to $\underline\Phi_2$ through Eq. (\ref{eq:  sitqmihf8}) should be understood in all the cases considered below.
\begin{itemize}
\item[(i)]A fixed unitary transformation (say $U$) applied to all operators (including the initial density operator). In this case, we have $\Phi_1$ = identity and $\Phi_2$ is the automorphism given by Eq. (\ref{eq: sitqmihf16}).
\item[(ii)] The same operator described in terms of fields at different times (using Heisenberg equations of motion), say at times $t$ and $t+a$. In this case, we have
\begin{eqnarray}
\label{eq: peoh1}		
& \Phi_1(t) = t+a; \hspace{2cm} & \Phi_{2t} = V(t+a,t)		
\end{eqnarray}				
where $V(t,t')$ is the evolution operator in our formalism (given by the $\tilde U$ of Eq. (\ref{eq: sitqmihf16}) corresponding to $U = U(t,t') = \exp [-iH(t-t')/\hbar]$).
\item[(iii)] Histories related through field redefinitions: Given a field transformation $(\phi,\pi)\rightarrow(\phi',\pi')$, the two histories involve, typically, observables $A$ and $A'$ related through equations like $A[\phi,\pi]=A'[\phi',\pi']$. Since, as operators, $A$ and $A'$ are identical (and have, therefore, identical spectral projectors), the mappings $\Phi_1$ and $\Phi_2$ are both identity maps. As an example, let $\phi' = U^{-1}\phi U$, $\pi'=U^{-1}\pi U$,  $A=\int\pi\dot{\phi}\,dx$; then 
\begin{equation}
\label{eq: peoh2}		
A[\phi,\pi]\equiv\int\pi\dot{\phi}\,dx = U [\int\pi'\phi'\,dx]U^{-1} \equiv A'[\phi',\pi'].
\end{equation}			
\item[(iv)] Another notion of physical equivalence is that two histories described, respectively, by triples $(\{C_\alpha\},H,\rho)$ and $(\{\tilde{C}_\alpha\},\tilde{H},\tilde{\rho})$ are physically equivalent if there exist canonical pairs $(\phi,\pi)$ and $(\tilde\phi,\tilde\pi)$ such that the relevant operators in one history have the same expressions in terms of $(\phi,\pi)$ as those in the other have in terms of $(\tilde\phi,\tilde\pi)$ [here $C_\alpha$ are operators of the type of Eq. (\ref{eq: intro3}), $H$ is the Hamiltonian and $\rho$ is the initial density operator]. Assuming as above, that the theory in question has a concrete expression for the decoherence functional $d(\alpha,\beta)$ and denoting the histories corresponding to the two triples above as $\alpha$ and $\tilde\alpha$ respectively, we must have 
\begin{equation}
\label{eq: peoh3}		
d(\tilde\alpha,\tilde\beta)=d(\alpha,\beta).
\end{equation}				
This is because, when the two sides are expressed in terms of the canonical pairs $(\phi,\pi)$ and $(\tilde\phi,\tilde\pi)$, there is nothing to mathematically distinguish the two sides (apart from some trivial relabelling). In this case, we have $\Phi_1$ = identity and $\Phi_2$ precisely the mapping given by $\alpha\rightarrow\tilde\alpha$.
\end{itemize}
Remarks: (1) Note that all the notions of physical equivalence treated above can be \emph{derived} from our criterion. More generally, any formalism in which histories are described in terms of fundamental entities like particles, fields, strings \emph{etc.}, similar notions of physical equivalence of histories can be derived by showing that the mappings $\Phi_1$, $\Phi_2$, and $\Phi_3$ involved in the situation correspond to an orthochronous symmetry.

\noindent(2) The criterion of physical equivalence of histories stated above is applicable to \emph{closed} systems only. In particular, it is not applicable to the history version of traditional quantum mechanics in which events refer to measurements by an external observer. Gell-Mann and Hartle have emphasized that the criterion for physical equivalence of histories have to be different for closed systems (ideally, the universe where an observer or measurement apparatus is a part of the system) and for the `approximate quantum mechanics of a measured subsystem' of the universe. The latter categories of theories (of which standard quantum mechanics is an example) have external observers which employ reference frames; two histories related through a nontrivial transformation within a reference frame (for example, a time translation, a space translation or a spatial rotation) are physically distinguishable.


\section{Conservation laws}
\label{sec: cs}

We next consider conservation laws in the present formalism. Since we have the concept of evolution defined in the formalism, it is natural to define conservation of an observable in terms of its preservation under evolution. To do this, however, we shall need to compare elements of ${\mathcal U}_\tau$'s for different $\tau$'s. It follows that a primary requirement for the definition of conservation laws is that ${\mathcal U}_\tau$'s for different $\tau$'s be isomorphic. We shall assume it to be the case in this section and identify ${\mathcal U}_\tau$'s for all $\tau\in{\mathcal N}({\mathcal T})$.

Given $\tau\lhd\tau'$, we say an observable $A\in{\mathcal O}(\tau)$ is conserved under the evolution $V(\tau',\tau):{\mathcal U}_\tau\rightarrow{\mathcal U}_{\tau'}$ if
\begin{equation}
\label{eq: cs1}		
V(\tau',\tau)(A(E)) = A(E) \hspace{2cm} \mbox{for all } E\in B(R).
\end{equation}				
We shall now formulate an alternative definition of conservation law in terms of equality of probabilities of appropriate single-time histories. Let $\tau_0\lhd\tau\lhd\tau'$ and suppose that a decoherence functional $d_{p_0,V}$ is given in terms of an initial state $p_0$ at `time' $\tau_0$ and the evolution map $V(.,.)$. Let $\alpha$ and $\beta$ be the single-time histories given by
\begin{eqnarray}
\label{eq: cs2}		
\alpha = \alpha_{\tau} = A(E); & \hspace{2cm} & \beta = \beta_{\tau'} = A(E).
\end{eqnarray}				
We expect the following alternative definition of conservation law to be equivalent to the one given: we say that the observable $A\in{\mathcal O}(\tau)$ is conserved under the evolution $V(\tau',\tau):{\mathcal U}_\tau\rightarrow{\mathcal U}_{\tau'}$ if
\begin{equation}
\label{eq: cs3}		
d_{p_0,V}(\alpha,\alpha) = d_{p_0,V}(\beta,\beta)
\end{equation}				
for all $p_0\in{\mathcal S}(\tau_0)$ and all $E\in B(R)$. A formal proof of this equivalence is possible only if either an explicit functional form of the decoherence functional is known or more detailed assumptions about its dependence on the evolution mapping $V$ are made. In the following, we shall only verify that the two definitions are indeed equivalent for history versions of nonrelativistic quantum mechanics and classical mechanics.
\begin{enumerate}				
\item[(i)] Quantum Mechanics: Let $t_0 < t < t'$. We have ${\mathcal U}_{t_0}\approx{\mathcal U}_{t}\approx{\mathcal U}_{t'}={\mathcal P}({\mathcal H})$. With the initial state $\rho(t_0)=\rho_0$, $\alpha=\alpha_t\in{\mathcal P}({\mathcal H})$ and $\beta=\beta_{t'}=\alpha_{t}$, we have [recalling Eqs. (\ref{eq: intro3}) and (\ref{eq: intro5})] $C_\alpha = \alpha_t U(t,t_0)$, $C_\beta = \beta_{t'} U(t',t_0)$ and
\alphaeqn
\begin{eqnarray}
\label{eq: cs4}		
d(\alpha,\alpha) & = &  \mbox{Tr}(C_\alpha\rho_0 C^{\dagger}_\alpha) =\mbox{Tr}\left[\alpha_t U(t,t_0)\rho_0 U(t,t_0)^{-1}\right] \\
d(\beta,\beta) & = & \mbox{Tr}\left[\beta_{t'}U(t',t_0)\rho_0 U(t',t_0)^{-1}\right]. 
\end{eqnarray}				
\reseteqn
The equality $d(\alpha,\alpha)=d(\beta,\beta)$ for arbitrary $\rho_0$ gives
\begin{equation}
\label{eq: cs5}		
U(t',t_0)^{-1}\beta_{t'} U(t,t_0) = U(t,t_0)^{-1} \alpha_t U(t,t_0)
\end{equation}				
which implies (recalling that $\beta_t'=\alpha_t$)
\begin{equation}
\label{eq: cs6}		
U(t',t)^{-1}\alpha_t U(t',t) = \alpha_t.
\end{equation}				
This equation represents the preservation of spectral projectors of the observable $A$ under Schr\"{o}dinger picture evolution. Consider, for example
\begin{displaymath}			
\alpha_t = |\Psi_t\rangle\langle\Psi_t|\in{\mathcal U}_t
\end{displaymath}			
where $|\Psi_t\rangle$ is an eigenstate of $A$ corresponding to some eigenvalue $\lambda$. Under temporal evolution $|\Psi_t\rangle \rightarrow U(t',t)|\Psi_t\rangle$ and
\begin{equation}
\label{eq: cs7}		
|\Psi_t\rangle\langle\Psi_t| \rightarrow U(t',t)|\Psi_t\rangle\langle\Psi_t|U(t',t)^{-1} \in{\mathcal U}_{t'}.
\end{equation}				
A general spectral projector of $A$ (which is a sum or integral of projectors of the form $|\Psi_t\rangle\langle\Psi_t|$) has the same transformation law under temporal evolution. The content of Eq. (\ref{eq: cs6}), therefore, is the same as that of Eq. (\ref{eq: cs1}) in the present case. The condition (\ref{eq: cs3}), therefore, implies condition (\ref{eq: cs1}). Conversely, given Eq. (\ref{eq: cs1}), we have Eq. (\ref{eq: cs6}) from which we can easily obtain the equality (\ref{eq: cs3}).
\item[(ii)] Classical Mechanics: Again, let $t_0 < t<t'$. We have ${\mathcal U}_{t_0}\approx{\mathcal U}_{t}\approx{\mathcal U}_{t'}= B(\mit\Gamma)$. With $p_0\in{\mathcal S}(t_0)$, $\alpha=\alpha_t=A(E)\in B(\mit\Gamma)$ and $\beta=\beta_{t'}=\alpha_t$ we have
\alphaeqn
\begin{eqnarray}		
d(\alpha,\alpha) & = & \int_{\mit\Gamma} dp_0(\xi_0)\int_{\alpha_t} \delta\left[\xi-V(t,t_0)(\xi_0)\right] d\xi \nonumber \\
 & = & p_0\left[V(t,t_0)^{-1}(\alpha_t)\right] \label{eq: cs8}. \\
d(\beta,\beta) &  =  & p_0\left[V(t',t_0)^{-1}(\beta_{t'})\right]. \label{eq: cs9}
\end{eqnarray}						
\reseteqn
The equality $d(\alpha,\alpha)=d(\beta,\beta)$ for all $p_0$ gives 
\begin{equation}
\label{eq: cs10}		
V(t,t_0)^{-1}(\alpha_t) = V(t',t_0)^{-1}(\beta_{t'}),
\end{equation}				
which implies $V(t',t)(A(E))=A(E)$ [Eq.~(\ref{eq: cs1})]. Conversely, given Eq. (\ref{eq: cs1}), we have Eq. (\ref{eq: cs10}) from which Eq. (\ref{eq: cs3}) is easily deduced.
\end{enumerate}
						
Our definition of a conservation law is somewhat different from the one given in~\cite{hlm1} where, in the limited context of Hilbert space quantum mechanics (of a closed system), conservation of an observable $A$ is defined in terms of vanishing probability of decoherent histories which involve projection operators corresponding to disjoint ranges of $A$ at two different times (making allowance for finite sequences of projectors associated with other observables at intermediate times). Our definition, in contrast, is given (in a more general framework) in terms of equality of probabilities of single-time histories at different `times' for all ranges of $A$ and for all initial states [see Eq. (\ref{eq: cs3})]. The definition in~\cite{hlm1} raises interesting questions about permitted projectors at intermediate times which are investigated there and some interesting results obtained. 


\section{Connection between symmetries and conservation laws in Hilbert space-based theories}
\label{sec: cbsacl}

In certain forms of dynamics, one can obtain a general relation between continuous symmetries of dynamics and conservation laws. Examples are Lagrangian and Hamiltonian formulations of classical mechanics and Hilbert space quantum mechanics in the Heisenberg picture. The conserved quantities are infinitesimal generators of symmetry transformations (in the appropriate implementation of symmetry) interpreted as observables. In the present formalism, it is difficult to see such a general connection between symmetries and conservation laws in the case of general ${\mathcal U}_\tau$'s. At least two conditions appear to be necessary to establish such a general connection: (i) An explicit expression for the decoherence functional $d_{p,V}$ [or at least some information about its dependence on the evolution maps $V(\tau',\tau)$]. (ii) The presence of an appropriate mathematical structure which can lead to the identification of the infinitesimal generator of a symmetry transformation as an observable. These conditions are satisfied for Hilbert space based theories in which case we derive such a general relation. Before doing that we quickly recall the definition of decoherence functionals in such theories.

 We are considering the subclass of theories described in section~\ref{sec: hbgqm} in which  we have, for each $\tau\in{\mathcal N}({\mathcal T})$, a separable Hilbert space ${\mathcal H}_\tau$ and ${\mathcal U}_\tau={\mathcal P}({\mathcal H}_\tau)$, the family of projection operators in ${\mathcal H}_\tau$. Here ${\mathcal T}$ is a general space of temporal supports satisfying axioms $A_1$ and $A_2$ of section~\ref{sec: hbgqm}. For $\tau,\tau'\in{\mathcal N}({\mathcal T})$ with $\tau\lhd\tau'$, the evolution from ${\mathcal U}_\tau$ to ${\mathcal U}_{\tau'}$ is, for the purpose at hand, more conveniently described by the linear map $K(\tau',\tau):{\mathcal H}_\tau\rightarrow{\mathcal H}_{\tau'}$ such that, for all triples $\tau,\tau',\tau''$ with $\tau\lhd\tau'\lhd\tau''$ and $\tau\lhd\tau''$, we have $K(\tau'',\tau')\cdot K(\tau',\tau) = K(\tau'',\tau)$. [When ${\mathcal H}_\tau$'s are naturally isomorphic and the maps $K(.,.)$ unitary, the transformation law $\alpha\rightarrow K(\tau',\tau)\alpha K(\tau',\tau)^\dagger$ of projectors gives the mappings $V(\tau,\tau'):{\mathcal P}({\mathcal H}_\tau)\rightarrow{\mathcal P}({\mathcal H}_{\tau'})$ which are logic isomorphisms.] We also assume that for every pair $\tau,\tau'\in{\mathcal N}({\mathcal T})$, there exists a $\tau''\in{\mathcal N}({\mathcal T})$ such that $\tau\lhd\tau''$ and $\tau'\lhd\tau''$. (This is a new assumption not covered by the axioms of section~\ref{sec: hbgqm}.) In the HPO formalism (section~\ref{subsec: hpof}), elements of $\tilde{\mathcal U}$ are represented as homogeneous projectors
\begin{equation}
\label{eq: cbsacl1}		
\tilde\alpha = \alpha_{\tau_1}\otimes\alpha_{\tau_2}\otimes\cdots\otimes
\alpha_{\tau_n}
\end{equation}				
where $\tau_1\lhd\tau_2\lhd\cdots\lhd\tau_n$ and $\alpha_{\tau_i}\in{\mathcal U}_{\tau_i} = {\mathcal P}({\mathcal H}_{\tau_i})$. Similarly, elements of $\Omega$ are represented as inhomogeneous projectors. Due to countable additivity of the decoherence functionals, it is sufficient to define $d(.,.)$ for pairs of homogeneous projectors. Writing $\tilde\alpha$ for $\{\tilde{\alpha}\}$ and taking 
\begin{displaymath}			
\tilde\beta = \beta_{\tau'_1}\otimes\cdots\otimes\beta_{\tau'_m}, 
\hspace{1cm} \tau'_1\lhd\cdots\lhd\tau'_m
\end{displaymath}			
we define [compare Eq. (\ref{eq: intro5})]
\begin{equation}
\label{eq: cbsacl2}		
d(\tilde\alpha,\tilde\beta) = N \mbox{Tr}\left[ C'_{\alpha}\rho(\tau_0) C^
{'\dagger}_\beta\right]		
\end{equation}				
where $\tau_0\lhd\tau_1$, $\tau_0\lhd\tau'_1$, $\rho(\tau_0)$ is a density operator on ${\mathcal H}_{\tau_0}$ and 
\begin{equation}
\label{eq: cbsacl3}		
C'_{\alpha} = K(\tau_f,\tau_n)\alpha_{\tau_n}K(\tau_n,\tau_{n-1})\cdots\alpha_{\tau_2}K(\tau_2,\tau_1)\alpha_{\tau_1}K(\tau_1,\tau_0)
\end{equation}				
with a similar expression for $C'_\beta$. Here $\tau_f$ is any element of ${\mathcal N}({\mathcal T})$ satisfying the conditions $\tau_n\lhd\tau_f$ and $\tau'_m\lhd\tau_f$. Moreover $N^{-1}= \mbox{Tr}\left[A\rho(\tau_0) B\right]$ where $A$ and $B$ are the operators obtained from $C'_\alpha$ and $C'_\beta$ respectively by putting each of the $\alpha_{\tau_i}$ and $\beta'_{\tau_j}$ equal to the unit operator.

Since conservation laws can be defined only when all ${\mathcal U}_\tau$'s are isomorphic, we restrict ourselves to the subclass of Hilbert space-based theories in which all the ${\mathcal H}_\tau$'s are naturally isomorphic (and can, therefore be identified with a single Hilbert space ${\mathcal H}$) and the evolution maps are unitary. This means one has the usual Hilbert space-based quantum mechanics with unitary temporal evolution except that the time points $t,t',\cdots$ are replaced by the nuclear elements of a psg. Even this much generality, however, is worthwhile; results obtained will have validity in, for example, quantum field theories in a large class of space-times not admitting foliation in spacelike surfaces.

We now proceed to obtain the desired relation between continuous symmetries and conservation laws.

Let $\Phi(\lambda)=(\Phi_1(\lambda),\Phi_2(\lambda),\Phi_3(\lambda))$ be a continuous 1-parameter symmetry of dynamics of a history system in the above mentioned class. Such a symmetry is implemented unitarily on ${\mathcal H}$ (section~\ref{subsec: rws});  let the corresponding infinitesimal generator be the self-adjoint operator $A$. The invariance condition (\ref{eq: siatlf9}) of $V$ [with transformation law (\ref{eq: siatlf6}) for $V(.,.)$] implies that the mappings $K(.,.)$ mentioned above commute with $A$ and therefore with the spectral projectors of $A$ (see Appendix \ref{app: two} for a proof).

Now let $\tau_0\lhd\tau\lhd\tau'\lhd\tau_f$, $\tau_0\lhd\tau_f$ and $\tau\lhd\tau_f$ and consider the quantities $d(\alpha,\alpha)$ and $d(\beta,\beta)$ with $d(.,.)$ of Eq. (\ref{eq: cbsacl2}). Here we have $N=1$ and $\alpha$, $\beta$ are single-time histories given by $\alpha_\tau=\beta_{\tau'}=P$, a spectral projector of $A$. We have, in this case,
\begin{eqnarray}
\label{eq: cbsacl4}		
C'_\alpha & = & K(\tau_f,\tau) P K(\tau,\tau_0) \nonumber \\
 & = & K(\tau_f,\tau) K(\tau,\tau_0) P = K(\tau_f,\tau_0) P.
\end{eqnarray}				
This gives
\begin{eqnarray}
\label{eq: cbsacl5}		
d(\alpha,\alpha) & = & \mbox{Tr}\left[ C'_\alpha \rho(\tau_0) C^{'\dagger}_\alpha \right] \nonumber \\
 & = & \mbox{Tr}\left[ K(\tau_f,\tau_0) P \rho(\tau_0) P K(\tau_f,\tau_0)^{\dagger}\right] \nonumber \\
 & = & \mbox{Tr}\left[P\rho(\tau_0)\right]. 	
\end{eqnarray}				
Similarly, we have $d(\beta,\beta) = \mbox{Tr}\left[P\rho(\tau_0)\right] = d(\alpha,\alpha)$ giving the desired conservation law in the form of Eq. (\ref{eq: cs3}).


\section{Histories in curved spacetimes}
\label{sec: hics}
In this section we consider histories in general curved spacetimes. As mentioned earlier, the treatment is general enough to cover histories of particles, fields, strings and more general objects.


\subsection{Quasitemporal structure}
\label{subsec: qs}

First we consider the construction of the space ${\mathcal T}$ of temporal supports. We assume that the spacetime $M$ is a 4-dimensional manifold equipped with a metric $g$ of Lorentzian signature $(- + + +)$. Given two subsets $A$ and $B$ of $M$, we say that $A$ temporally precedes $B$ $(A\lhd B)$ if 
\begin{eqnarray}
\label{eq: qs1}
J^+(A)\cap B\neq\phi, & \hspace{1cm} & \left[ J^+(B)-B\right]\cap A =\phi.
\end{eqnarray}
We have modified Isham's definition (\ref{eq: intro14}) to allow the possibility $A = B$ or, more generally, $A\cap B\neq\phi$. This is in keeping with our convention (\ref{eq: intro10}) which allows, for example, $s_m = t_1$ in Eq. (\ref{eq: intro9}). Like the relation $\prec$ given by Eq. (\ref{eq: intro14}), the relation $\lhd$ is also not a partial order; in particular, $A\lhd B$ and $B\lhd C$ does not generally imply $A\lhd C$.

We define a {\it basic region} as a connected subset of $M$ such that every pair of points in it has spacelike separation. Thus, a basic region can be a single point, (a connected piece of) a spacelike curve, a two-dimensional spacelike surface or a three dimensional spacelike hypersurface. This is in contrast to Isham's definition where the basic regions are four dimensional.

A nuclear temporal support is a collection
\begin{equation}
\label{eq: qs2}
\tau=\{B_1,B_2,\ldots\}
\end{equation}
of basic regions such that all the pairs $B_i$, $B_j$ have mutually spacelike separation. Given two nuclear temporal supports $\tau=\{B_1,B_2,\ldots\}$ and $\tau'= \{B'_1,B'_2,\ldots\}$, we say that $\tau\lhd\tau'$ if 
\begin{equation}
\label{eq: qs3}
\cup_i(B_i) \lhd \cup_j(B'_j).
\end{equation}
A temporal support
\begin{equation}
\label{eq: qs4}
\xi=\{\ldots,\tau_1,\tau_2,\ldots\}
\end{equation}
is a countable ordered collection of nuclear temporal supports such that consecutive entries in $\xi$ are temporally ordered (\emph{i. e.} $\tau_j\lhd\tau_{j+1}$ for all $j$), and moreover, the collection is at most semi-infinite (it has either an `earliest' member or a `latest' member or both). The family of temporal supports is denoted as ${\mathcal T}$ and the subfamily of nuclear temporal supports as ${\mathcal N}({\mathcal T})$. Identifying $\tau\in{\mathcal N}({\mathcal T})$ with $\{\tau\}\in{\mathcal T}$, ${\mathcal N}({\mathcal T})$ is clearly a subfamily of ${\mathcal T}$.

We define a partial semigroup (psg) structure on ${\mathcal T}$ as follows: Given two elements $\xi$ and $\eta$ of ${\mathcal T}$, we say that $\xi\lhd\eta$ if $\xi$ has a latest member $\tau_0$ and $\eta$ has an earliest member $\tau'_1$ (\emph{i. e.} $\xi=\{\ldots,\tau_{-1},\tau_0\}$ and $\eta=\{\tau'_{1},\tau'_{2},\ldots\}$) such that $\tau_0\lhd\tau'_1$. If the joint collection of nuclear temporal supports in $\xi$ and $\eta$ is at most semi-infinite, we define the composition $\xi\circ\eta$ as 
\begin{equation}
\label{eq: qs5}
\xi\circ\eta=\{\ldots,\tau_{-1},\tau_0,\tau'_1,\tau'_2,\ldots\}.
\end{equation}
We assume that [see Eq. (\ref{eq: intro10})]
\begin{equation}
\label{eq: qs6}
\{\tau\}\circ\{\tau\}=\{\tau\} \mbox{    for all }\tau\in{\mathcal N}({\mathcal T}).
\end{equation}
The composition (\ref{eq: qs5}) is  now also defined if $\tau_0=\tau'_1$.

A general element $\xi=\{\ldots,\tau_1,\tau_2,\ldots\}\in{\mathcal T}$ admits an irreducible decomposition of the form
\alphaeqn 
\begin{equation}
\label{eq: qs7}  
\xi=\cdots\circ\{\tau_1\}\circ\{\tau_2\}\circ\cdots
\end{equation}
which we simply write as
\begin{equation}
\label{eq: qs8}
\xi=\cdots\circ\tau_1\circ\tau_2\circ\cdots.
\end{equation}
\reseteqn
This irreducible decomposition is unique modulo the trivial redundancy implied by the convention (\ref{eq: qs6}).

In the appendix~\ref{app: one} we give a straightforward construction of a topology on ${\mathcal T}$ to make it a topological partial semigroup (\emph{i. e.} a psg ${\mathcal T}$ which is also a topological space with a topology such that the composition $\circ$, considered as a mapping from a subset of ${\mathcal T}\times{\mathcal T}$ into ${\mathcal T}$ is continuous). This is needed for a continuity argument in the next subsection.

We next consider the construction of the space ${\mathcal U}$ of history filters. The first step in this construction is to associate, with each basic region $B$, a logic ${\mathcal U}_B$ such that, for two basic regions $B$ and $B'$ having spacelike separation, the logics ${\mathcal U}_B$ and ${\mathcal U}_{B'}$ are isomorphic. Identifying isomorphic logics, we can now associate a logic ${\mathcal U}_\tau$ with an element $\tau\in{\mathcal N}({\mathcal T})$. (It is just the logic associated with any of its basic regions.) Logics associated with two different nuclear temporal supports generally need not be isomorphic. 

A history filter is now defined as an assignment, to some element $\xi=\{\ldots,\tau_1,\tau_2,\ldots\}$ of ${\mathcal T}$, of a collection $\alpha=\{\cdots,\alpha_1,\alpha_2,\cdots\}$ such that
\begin{itemize}
\item[(i)] the entries in $\alpha$ are in one-one correspondence with those in $\xi$;
\item[(ii)] $\alpha_j\in{\mathcal U}_{\tau_j}$ for every $j$.
\end{itemize}
We define the map $\sigma:{\mathcal U}\rightarrow{\mathcal T}$ such that $\sigma(\alpha)=\xi$. A temporal order relation $\lhd$ and a psg structure can now be defined on ${\mathcal U}$ in a fairy obvious manner so as to make $\sigma$ a psg homomorphism. Indeed, given $\alpha=\{\cdots,\alpha_{-1},\alpha_0\}$ and $\beta=\{\beta_1,\beta_2,\cdots\}$ with $\sigma(\alpha)=\xi=\{\ldots,\tau_{-1},\tau_0\}$ and $\sigma(\beta)=\eta=\{\tau'_1,\tau'_2,\ldots\}$ we say that $\alpha\lhd\beta$ if $\xi\lhd\eta$ (which means $\tau_0\lhd\tau'_1$). Moreover, if $\xi\circ\eta$ is defined, we define
\begin{equation}
\label{eq: qs9}
\alpha\circ\beta=\{\cdots,\alpha_{-1},\alpha_0,\beta_1,\beta_2,\cdots\}.
\end{equation}
Clearly
\begin{eqnarray}
\label{eq: qs10}
\sigma(\alpha\circ\beta)= & \{\ldots,\tau_{-1},\tau_0,\tau'_1,\tau'_2,\ldots\}= \xi\circ\eta &=\sigma(\alpha)\circ\sigma(\beta).
\end{eqnarray}
In keeping with the convention (\ref{eq: qs6}) for ${\mathcal T}$, we adopt a similar convention for the space ${\mathcal U}$: Given $\alpha={\alpha_1}$ such that $\sigma(\alpha)={\tau_1}$ where $\tau_1\in{\mathcal N}({\mathcal T})$, we stipulate that 
\begin{equation}
\label{eq: qs11}
\{\alpha_1\}\circ \{\alpha_1\}=\{\alpha_1\}.
\end{equation}
The mapping $\sigma:{\mathcal U}\rightarrow{\mathcal T}$ is a psg homomorphism  and the psg's  ${\mathcal U}$ and ${\mathcal T}$ are directed and special. The triple $({\mathcal U},{\mathcal T},\sigma)$ satisfies the axiom $A_1$ of section~\ref{sec: hbgqm}. We can now invoke the other axioms, and, proceeding as in section~\ref{sec: hbgqm}, construct the space $\Omega$ of history propositions (`inhomogeneous histories') which is manifestly an orthoalgebra~\cite{dj1}. 

It should be mentioned that the axiom $A_2$ of section~\ref{sec: hbgqm}, which excludes histories with `closed time loops' would exclude spacetimes admitting closed timelike curves.

The logics employed above can be quite general - they can be Boolean logics associated with classical mechanics of particles and/or fields, standard quantum logics (typically the space ${\mathcal P}({\mathcal H})$ of projection operators in a separable Hilbert space ${\mathcal H}$) or more general logics. The formalism presented above can, therefore, be applied to the history versions of the classical or quantum mechanics of particles, fields, strings or more general objects.

Explicit construction of the decoherence functionals for the two special subclasses of theories mentioned in the introduction can be easily adopted in the corresponding subclasses of theories considered in the present section (\emph{i. e.} when ${\mathcal U}_\tau$'s are either families of projection operators in separable Hilbert spaces or those of Borel measurable subsets of phase spaces of classical systems); we shall, however, skip the details.


\subsection{Symmetries}
\label{subsec: s}
The definition of symmetry given in section~\ref{subsec: siatlf} is quite general and can be adopted in the formalism of previous subsection. The mapping which has some special features in the present context is $\Phi_1$. (Special features will emerge in $\Phi_2$ if more structure in the space ${\mathcal U}$ is incorporated in terms of, for example, field theoretic notions. This will not be done here.) We shall, therefore, concentrate on the mapping $\Phi_1$ in the remainder of this section. 

The mapping $\Phi_1$, being an (anti-) automorphism of the partial semigroup ${\mathcal T}$, maps nuclear elements to nuclear elements in a one-to-one manner. Recall that the nuclear elements of ${\mathcal T}$ are defined as collections of basic regions [see Eq. (\ref{eq: qs2})] and that our definition allows points, spacelike curves and 2- and 3-dimensional spacelike regions to be basic regions. All the basic regions are, of course, special cases of nuclear temporal supports. Now, any mapping between two sets preserves inclusion relations between their subsets. It follows that $\Phi_1$ maps, in one-to-one manner, basic regions onto basic regions preserving their dimensionalities. It follows, in particular, that, considering $M$ as a subset of ${\mathcal T}$, $\Phi_1$ induces an invertible mapping of $M$ onto itself (which we shall denote as $\Phi^M_1$). Symmetries for which $\Phi_1^M$ is the identity mapping may be called internal symmetries. (We have, therefore, a way of defining internal symmetries in this formalism.)

We prove below that $\Phi^M_1$ must be a conformal isometry of the spacetime $(M,g)$. In the proof, we need the continuity of mapping $\Phi^M_1$. To facilitate such an argument, we employ the topology on ${\mathcal T}$ constructed in Appendix~\ref{app: one} which makes it a topological partial semigroup. We impose the continuity condition on the mapping $\Phi_1$ (and show in appendix~\ref{app: one} that (considering $M$ as a subset of ${\mathcal T}$), the subspace topology on $M$ coincides with the manifold topology of $M$. Hence the continuity of $\Phi_1$ implies continuity of $\Phi^M_1$ in the manifold topology. [Similar continuity requirements should be understood on $\Phi_2$ and $\Phi_3$ (with appropriate topologies defined on relevant spaces); they will, however, not be discussed here.] 

We, therefore, have an invertible continuous mapping $\Phi^M_1:M\rightarrow M$ which maps spacelike curves onto spacelike curves. It follows that, it maps non-spacelike curves onto non-spacelike curves. Now, null curves can be realized as limits of spacelike curves. Being continuous, $\Phi^M_1$ must map null curves onto null curves and therefore timelike curves onto timelike curves. Considering the transformation of a small neighborhood of a point $p$ of $M$ and employing the usual local coordinate representation of line elements, we have, therefore, 
\begin{equation}
\label{eq: s5}
g'_{\mu\nu}(p')\Delta {x^\mu}^{'} \Delta {x^\nu}^{'} = \lambda_p g_{\mu\nu}(p)\Delta x^\mu \Delta x^\nu 
\end{equation}
where primes indicate transformation under $\Phi_1$ and $\lambda_p$ is a positive constant (possibly dependent on $p$). Since the point $p$ in Eq. (\ref{eq: s5}) is arbitrary, the mapping $\Phi^M_1$ must be a conformal isometry of the spacetime $(M,g)$.

In the present context, orthochronous symmetries are those preserving the temporal order of the basic regions. This implies, on considering a light cone as a disjoint union of basic regions with temporal relations between them, that, for an orthochronous symmetry, the mapping $\Phi_1^M$ maps past lightcones of spacetime points to past lightcones and future lightcones to future lightcones. In contrast, the non-orthochronous symmetries reverse the temporal order of basic regions;  consequently, they map past lightcones to future lightcones and {\it vice versa}.


\section{Concluding remarks}
\label{sec: cr}

The main inadequacy which has seriously affected the present work is the absence of a concrete expression for the decoherence functional $d_{p,V}$ in the general case. Construction of such a functional is an important problem which deserves serious effort at solution. There have been several attempts in literature~\cite{jh2,il1,ils1,jdmw1} at construction of decoherence functional in various situations and at obtaining some general results about decoherence functionals~\cite{ss3,il2}. These, however, do not appear to be adequate to solve the above mentioned problem. 

Even in the absence of such an expression, we have shown in this work that a straightforward definition of symmetry can be given (which can be easily adapted to situations when a concrete expression for $d_{p,V}$ is available) which leads to some interesting results. An example of such a result obtained without a concrete decoherence functional is the formulation of a general criterion for physical equivalence of histories which covers the various notions of physical equivalence of histories considered by Gell-Mann and Hartle~\cite{gh1} as special cases (section~\ref{sec: peoh}). Examples of the results obtained using concrete decoherence functionals appeared in sections~\ref{sec: sinqm},~\ref{sec: sihvocm},~\ref{sec: cs} and~\ref{sec: cbsacl}.

Our treatment of quasitemporal structure for histories relating to dynamics of (closed) systems in curved spacetime is, although essentially along the lines of Isham~\cite{cji1} (in the sense that it employs a partial semigroup structure defined in terms of light cones), differs from it substantially in detail. Our definition [Eq. (\ref{eq: qs1})] of temporal order differs from that of Isham [Eq. (\ref{eq: intro14})] to achieve consistency with the conventions of~\cite{dj1}. [See Eq. (\ref{eq: intro10})]. More importantly, our basic regions are spacelike regions with dimensionality less than or equal to three in contrast to Isham's four-dimensional basic regions. Apart from the relative merits of the two schemes mentioned in section~\ref{sec: intro}, allowing basic regions of all dimensions less than or equal to three has the advantage that one can, for example, consider histories of systems involving both fields and particles. 

As a bonus, this generality has made it possible to obtain the interesting result (in section~\ref{sec: hics}) that a symmetry of a history theory relating to dynamics of systems (particles, fields, $\ldots$) in general curved spacetimes must have associated with it a transformation of the spacetime $M$ which is a conformal isometry of the underlying spacetime metric. Indeed all fundamental symmetries in various domains of physics - nonrelativistic/relativistic, classical/quantum, particle/field/string dynamics - satisfy this condition. It is, indeed, very satisfying that such a result should appear at the present level of generality.

In a relatively limited context, a theorem of this type is the generalized Wigner theorem discussed in section~\ref{subsec: siha}; it was obtained in Houtappel {\it et al.}~\cite{hvw1} using the $\Pi$-functions. It should be noted that this theorem characterizes the general symmetries in nonrelativistic quantum mechanics much more comprehensively than the traditional Wigner theorem does.

Before closing, we would like to stress upon an aspect of the histories-based theories which is not adequately appreciated in the literature. This relates to the fact that these are theories formulated in terms of the primitive elements of physical theory. Doing a little more systematic job of identifying these primitive elements than was done in~\cite{hvw1}, we propose these elements to be: (1) observations/observables, (2) evolution of systems, and (3) conditional predictions about systems. These are essentially the ingredients that go into the construction of histories. This aspect makes histories potentially useful in any domain of scientific study where physics is involved at the fundamental level - the problem of scientific explanation of consciousness~\cite{rp1,pmc1,djc1}. As an example of a work referring histories in such a context, see~\cite{cjsc1}. It is also worth mentioning here that, whereas the histories employed in the literature during the past one and a half decade refer to closed systems (for example, the universe), this was not the case for the $\Pi$-functions employed in~\cite{hvw1} where all measurements referred to external observers. It follows that the restriction to closed systems is not an  absolute requirement for viable history-based theories. This fact is of special significance in connection with the problem of consciousness where the systems of interest - the brain, the nervous system, the animal body - are by no means closed systems. 

\section*{Acknowledgements}
This work was supported, in part, by NSF grant no. DMR9714055.


\appendix
\section{Proof of the commutation of $K(.,.)$ with $A$.}
\label{app: two}

We give here the proof, needed in section~\ref{sec: cbsacl}, of the commutation of the evolution map $K( , )$ with the infinitesimal generator $A$ of the symmetry operation.

Eq. (\ref{eq: siatlf6}), along with the invariance condition (\ref{eq: siatlf9}) implies 
\begin{equation}
V(\tau'_{2},\tau'_{1})\circ\Phi_{2\tau_1}=\Phi_{2\tau_2}\circ V(\tau_2,\tau_1).
\label{eq: twoone}
\end{equation}
Since all the ${\cal U}_\tau$'s have been assumed isomorphic, we have $\Phi_{2\tau_1}=\Phi_{2\tau_2}$. The action of the maps $\Phi_{2\tau}$ and $V(\tau_2,\tau_1)$ on a projector $P$ is given, in terms of the objects $U(\lambda)$ and $K(\tau_2,\tau_1)$, introduced in section~\ref{sec: cbsacl}, by
\begin{eqnarray}
\Phi_{2\tau} (P) &  = & U(\lambda) P U(\lambda)^{\dagger}, \\ \label{eq: twotwo} 
V(\tau_2,\tau_1) (P) & = & K(\tau_2,\tau_1) P K(\tau_2,\tau_1)^{\dagger} \label{eq: twothree}.
\end{eqnarray}
Applying the mappings appearing on the two sides of Eq. (\ref{eq: twoone}) to the projector $P = |\psi\rangle\langle\psi|$, we get
\begin{equation}
|\psi_1\rangle\langle\psi_1|=|\psi_2\rangle\langle\psi_2|
\label{eq: twofour}
\end{equation}
where 
\begin{eqnarray}
|\psi_1\rangle = K(\tau'_2,\tau'_1)U(\lambda)|\psi\rangle, & \hspace{1cm} & |\psi_2\rangle =U(\lambda)K(\tau_2,\tau_1)|\psi\rangle \nonumber. 
\end{eqnarray}
Putting $\langle\psi_2|\psi_1\rangle = a$ (note that `$a$' is a function of $|\psi\rangle$ and some other quantities), Eq. (\ref{eq: twofour}) gives
\begin{equation}
|\psi_1\rangle = a |\psi_2\rangle = |a|^2 |\psi_1\rangle
\label{eq: twofive}
\end{equation}
which gives $|a|=1$. The first relation in (\ref{eq: twofive}) gives 
\begin{equation}
K(\tau'_2,\tau'_1) U(\lambda)|\psi\rangle = a U(\lambda)K(\tau_2,\tau_1)|\psi\rangle.
\label{eq: twosix}
\end{equation}
Taking, for an infinitesimal transformation, $U(\lambda)= I+\lambda A$, Eq. (\ref{eq: twosix}) gives the relations
\begin{eqnarray}
K(\tau'_2,\tau'_1) |\psi\rangle &  = & a K(\tau_2,\tau_1)|\psi\rangle, \\ \label{eq: twoseven}
K(\tau'_2,\tau'_1) A |\psi\rangle &  = & a  A K(\tau_2,\tau_1)|\psi\rangle \label{eq: twoeight}
\end{eqnarray}
which, in turn, give
\begin{equation}
K(\tau'_2,\tau'_1) A |\psi\rangle = A K(\tau'_2,\tau'_1)|\psi\rangle.
\label{eq: twonine}
\end{equation}
Since $|\psi\rangle$ is arbitrary, this gives the desired relation
\begin{equation}
K(\tau'_2,\tau'_1) A = A K(\tau'_2,\tau'_1).
\label{eq: twoten}
\end{equation}


\section{Construction of topology for the space of temporal supports constructed in section IX}
\label{app: one}

First we recall some definitions and results about topological spaces~\cite{jlk1}. We shall assume that the reader is familiar with the definition of topology in terms of open sets, neighborhoods and the concept of continuity of mappings between topological spaces.

Let $(X,{\mathcal O})$ be a topological space (which means that $X$ is a nonempty set and ${\mathcal O}$ is the family of open subsets of $X$). A family ${\mathcal B}$ of subsets of $X$ is a \emph{base} for the topology ${\mathcal O}$ if, for each point $x$ of $X$ and each neighborhood $V$ of $x$, there is a member $W$ of ${\mathcal B}$ such that $x\in W\subset V$. If ${\mathcal B}$ is a subfamily of ${\mathcal O}$ , it is a base for ${\mathcal O}$ if and only if each member of ${\mathcal O}$ is a union of members of ${\mathcal B}$.

A family ${\mathcal S}$ of subsets of $X$ is a \emph{subbase} for the topology ${\mathcal O}$ if the family of finite intersections of members of ${\mathcal S}$ is a base for ${\mathcal O}$ (equivalently, if each member of ${\mathcal O}$ is the union of finite intersections of members of ${\mathcal S}$). Every nonempty family ${\mathcal S}$ of subsets of a nonempty set $X$ is a subbase for {\it some} topology on $X$. This topology is {\it uniquely} defined by ${\mathcal S}$ and every member of ${\mathcal S}$ is an open set in the topology determined by ${\mathcal S}$.

A mapping $f$ of a topological space $X$ into a topological space $Y$ is continuous if and only if the inverse image under $f$ of every member of subbase for the topology on $Y$ is an open set in $X$. 

We shall construct a topology for ${\mathcal T}$ by choosing a family of subsets of ${\mathcal T}$ as a subbase. Given an element $\xi$ of ${\mathcal T}$ as in Eq. (\ref{eq: qs4}) with nuclear temporal supports $\tau_i$ of the form 
[see Eq. (\ref{eq: qs2})]
\begin{equation}
\label{eq: one1}
\tau_i=\{B_1^i,B_2^i,\ldots\}
\end{equation}
we introduce the collections (families of subsets of $M$)
\begin{equation}
\label{eq: one2}
N(\tau_i)=\{N_1(B_1^i),N_2(B_2^i),\ldots\}
\end{equation}
where $N_j(B_k^i)$ is an open neighborhood of the basic region $B_k^i$ in the manifold topology of $M$; we also introduce the collection
\begin{equation}
\label{eq: one3}
\tilde{N}(\xi) =\{\ldots,N_1(\tau_1),N_2(\tau_2),\ldots\}
\end{equation}
where each entry on the right is a collection of the form of Eq. (\ref{eq: one2}). Finally, we consider the family ${\mathcal F}$ of collections of the form (\ref{eq: one3}) for all the elements of ${\mathcal T}$: 
\begin{equation}
\label{eq: one4}
{\mathcal F}=\{\tilde{N}(\xi); \xi\in{\mathcal T}\}.
\end{equation}
We give a topology to the space ${\mathcal T}$ by stipulating that ${\mathcal F}$ be a subbase of that topology.

To show that, with this topology, ${\mathcal T}$ is a partial semigroup, we must show that the composition rule in ${\mathcal T}$ represented by an equation of the form $\xi\circ\eta=\zeta$ is a continuous mapping of (a subset of) the Cartesian product ${\mathcal T}\times{\mathcal T}$ into ${\mathcal T}$. To show this, it is adequate to show that the inverse image of any member of a subbase in the image space ${\mathcal T}$ is an open set in the product topology of ${\mathcal T}\times{\mathcal T}$. This is easily verified by making use of the definition of the composition rule in Eq. (\ref{eq: qs5}), the construction of the topology for ${\mathcal T}$ above and the definition of the product topology.

Since points of $M$ are basic regions and, therefore, (nuclear) elements of ${\mathcal T}$, $M$ can be considered as a subset of ${\mathcal T}$. We shall now show that the subspace topology of $M$ induced by the topology of ${\mathcal T}$ constructed above coincides with the manifold topology of $M$.

Open sets of $M$ in the subspace topology are the intersections of open sets of ${\mathcal T}$ with $M$. Consider first the family ${\mathcal F}_M$ consisting of intersections of the members of the family ${\mathcal F}$ with $M$. The members of ${\mathcal F}_M$ are subsets of $M$ which are open neighborhoods of points of $M$ in the manifold topology. These are also open subsets of $M$ in the subspace topology. Now, recalling the set theoretic relations
\begin{eqnarray}
\label{eq: one5}
\left(A\cap B\right)\cap M & = & \left(A\cap M\right)\cap\left(B\cap M
\right) \nonumber \\
\left(\cup_i B_i\right)\cap M & = & \cup_i\left(B_i\cap M\right) \nonumber
\end{eqnarray}
we see that the family ${\mathcal F}_M$ constitutes a subbase for the subspace topology of $M$. It is clearly also a subbase for the manifold topology of $M$. It follows that the subspace topology of $M$ coincides with its manifold topology. The choices in the initial steps in the construction of the topology for ${\mathcal T}$ above [see Eq. (\ref{eq: one2}) above] were made precisely to ensure this.


 
\end{document}